\documentclass[useAMS,usenatbib,a4]{mn2e}
\usepackage{times}
\usepackage{graphicx,bm,amssymb}
\usepackage{ulem}
\usepackage{epsfig}
\usepackage{amssymb}
\usepackage{natbib}
\usepackage{psfrag}
\usepackage[colorlinks=true,linkcolor=blue,citecolor=blue]{hyperref}
\def\apj{ApJ}%% Astrophysical Journal
\def\apjl{ApJ}%% Astrophysical Journal, Letters
\def\aap{A\&A}%% Astronomy and Astrophysics
\def\jcap{J. Cosmology Astropart. Phys.}%% Journal of Cosmology and Astroparticle Physics
\def\mnras{MNRAS}%% Monthly Notices of the RAS
\def\prd{Phys.~Rev.~D}%% Physical Review D
\def\nat{Nature}%% Nature
\def\physrep{Phys.~Rep.}%% Physics Reports
\def\be{\begin{equation}}
\def\ee{\end{equation}}
\def\bary{\begin{eqnarray}}
\def\eary{\end{eqnarray}}

\def\bi{\begin{itemize}}
\def\ei{\end{itemize}}
\def\lsim{\mathrel{\rlap{\lower3pt\hbox{\hskip1pt$\sim$}}
     \raise1pt\hbox{$<$}}} %less than or approx. symbol
\def\gsim{\mathrel{\rlap{\lower3pt\hbox{\hskip1pt$\sim$}}
     \raise1pt\hbox{$>$}}} %greater than or approx. symbol

\begin{document}
\title[Resonant oscillations of GeV - TeV neutrinos in internal shocks  from  gamma-ray burst  jets inside the stars]
{Resonant oscillations of GeV - TeV neutrinos in internal shocks  from  gamma-ray burst  jets inside the stars}
% Force line breaks with \\
\author[N. Fraija]%
{Nissim Fraija \thanks{E-mail:nifraija@astro.unam.mx }\\
Instituto de Astronom\' ia, Universidad Nacional Aut\'onoma de M\'exico, Circuito Exterior, \\C.U., A. Postal 70-264, 04510 M\'exico D.F., M\'exico}
%\date{\today} % It is always \today, today,

\maketitle
	
\begin{abstract}
High-energy neutrinos generated in collimated jets inside the progenitors of gamma-ray bursts (GRBs) have been related with the events detected by IceCube.  These neutrinos, produced by hadronic interactions of Fermi-accelerated protons with thermal photons and hadrons in internal shocks, are the only signature when jet has not broken out or failed.  Taking into account that the photon field is thermalized at keV energies and  the standard assumption that the magnetic field maintains a steady value throughout the shock region (with a width of  $10^{10} - 10^{11}$ cm in the observed frame),  we study the effect of thermal and magnetized plasma generated in internal shocks on the neutrino oscillations. We calculate the neutrino effective potential generated by this plasma, the effects of the envelope of the star, and the vacuum on the path to Earth.  By considering these three effects, the two (solar, atmospheric and accelerator parameters) and three neutrino mixing,  we show that although GeV - TeV neutrinos can oscillate resonantly from one flavor to another, a nonsignificant deviation of the standard flavor ratio (1:1:1) could be expected on Earth.
%
  %PACS numbers may be entered using the \verb+\pacs{#1}+ command.
\end{abstract}

%\pacs{98.70.Rz; 98.70.Sa}% PACS, the Physics and Astronomy
%\maketitle

\section{Introduction}\label{sec-Intro}
Long gamma-ray bursts (lGRBs) have been associated  to core collapse of massive stars leading to supernovae (CCSNe) of type Ib,c and II.  Type Ic supernovae are believed to be He stars with radius $R_\star\approx $ 10$^{11}$ cm, and type II and Ib are thought to have a radius of $R_\star\approx 3\times 10^{12}$ cm.  Depending on the luminosities and durations, successful lGRBs have revealed a variety of GRB populations: low-luminosity (ll),  ultra-long (ul) and high-luminosity (hl) GRBs \citep{PhysRevLett.87.171102,0004-637X-662-2-1111, 0004-637X-766-1-30}. While llGRBs and ulGRBs have a typical duration of ($\sim$ 10$^3$ - 10$^4$ s), hlGRBs have a duration of tens to hundreds of seconds. Another important population associated with CCSNe, although unobservables in photons, are failed  GRBs which could be much more frequent than successful ones, limited only by the ratio of type Ib/c and type II SNe to GRBs rates.  This population has been  characterized by having high-luminosities, mildly relativistic jets and durations from several to ten seconds \citep{2002MNRAS.332..735H, PhysRevLett.87.171102,2006Natur.442.1014S, 2010Natur.463..513S}.\\
Neutrinos are useful for studying the insides of stars, especially where photons cannot be observed either because jet fails or has not broken out yet, so in this case, they could be the only signature that would display the dynamics of the star.  High-energy (HE) neutrinos from this population of stars have been pointed out to contribute significantly to the extragalactic neutrino background (ENB) \citep{2013PhRvL.111l1102M, 2014MNRAS.437.2187F,tab10,2014arXiv1403.4089M,2013arXiv1312.0558W,2013PhRvD..88h1302R,2013PhRvL.111m1102M} and to explain the recent detections of TeV- PeV neutrinos by IceCube \citep{aar13, 2014arXiv1405.5303A}.\\
Measurements of  HE neutrino  properties such as  flavor content would be involved with new physics if a deviation of the standard flavor ratio were observed  \citep{lea95,ath00, kas05,2014arXiv1404.0017M}.  The neutrino flavor ratio  is expected to be at the source,  $\phi^0_{\nu_e}:\phi^0_{\nu_\mu}:\phi^0_{\nu_\tau}$=1 : 2 : 0 and on Earth (due to neutrino oscillations between the source and Earth) $\phi^0_{\nu_e}:\phi^0_{\nu_\mu}:\phi^0_{\nu_\tau}$=1 : 1 : 1 and  $\phi^0_{\nu_e}:\phi^0_{\nu_\mu}:\phi^0_{\nu_\tau}$=1 : 1.8 : 1.8  for neutrino energies lesser and greater than 300 TeV, respectively \citep{kas05}.  Also measurement of a non-zero $\theta_{13}$ mixing angle coming from astrophysical sources could be relevant to clarify the neutrino mass hierarchy as well as CP violation searches in neutrino oscillations \citep{2008PrPNP..60..338N, 2009RPPh...72j6201B,PhysRevD.86.073012}.\\
As known, neutrino properties are modified when they propagate in a thermal and magnetized medium.  A massless neutrino acquires an effective mass and an  effective potential.  The resonant conversion of active neutrino from one flavor to another  ($\nu_e \leftrightarrow \nu_\mu, \nu_\tau$) due to thermal and magnetized  medium has been explored in many astrophysical contexts and has had relevant consequences in the dynamics of them \citep{1978PhRvD..17.2369W, 1999A&A...344..573R, 1987ApJ...314L...7G, 2000APh....13...21V, 2008PhRvD..78c3014D, 2000PhLB..494..262E, 1996IJMPA..11..141D, 1994NuPhS..35..466D, 1996PhLB..383...87D, 1992PhRvD..46.1172D,  2003PhRvD..67b5018D, 1988NuPhB.307..924N, 1991NuPhB.349..754E}. For instance, \citet{2014ApJ...787..140F} has showed that the effect of magnetic field in the dynamics of the fireball evolution of GRB was to decrease the proton-to-neutron ratio aside from the number of multi-GeV neutrinos expected in a neutrino  detector.\\
Neutrino oscillations in vacuum and by matter effects in the failed GRB framework (along the jet and envelope of the star) have been examined by many authors \citep{men07, raz10, sah10, 2013arXiv1304.4906O, 2014MNRAS.437.2187F} and although these authors have studied the oscillations on the surface of the star due to its envelope, the effect of thermal and magnetic field plasma generated on internal shocks has not been explored.  In this paper we  calculate the effect of the magnetized and thermal shocked plasma on neutrino oscillations and then we estimate the flavor ratio on Earth. The organization of the paper is as follows: In section 2, we show a brief description of internal shocks.  In section 3, we derive firstly, the neutrino effective potential for  $m_W \leq E_\nu$  as a function of the magnetic field, temperature, angle (between the neutrino propagation and magnetic field)  and chemical potential and secondly,  the neutrino effective potential produced by the envelope of the star.  In section 4 we derive the resonance condition, the flip probability for two and three-neutrino mixing and the flavor ratio expected on Earth, and  in section V we discuss our results. We hereafter use $Q_x\equiv Q/10^x$ in c.g.s. units and k=$\hbar$=c=1 in natural units.
\section{Description of Internal shocks}
%The needed magnetic field is generated and amplified by the shocks 
%
One of the most prosperous theory to explain the prompt emission and the afterglow  in successful GRBs is the fireball model \citep{2004IJMPA..19.2385Z, 2006RPPh...69.2259M}.  A GRB is considered successful when the jet drills inside the progenitor and breaks through the stellar envelope, otherwise it is taken into account as a failed GRB.   When the jet encounters the stellar envelope two shocks are involved: an outgoing, or forward, shock \citep{1994ApJ...430L..93R, 1993ApJ...418L...5P} and another one that propagates back decelerating the ejecta, the reverse shock \citep{1994MNRAS.269L..41M, 1994ApJ...430L..93R}.  The jet dynamics is mainly dominated by the jet head, which is controlled by the ram pressure balance between the reverse  and forward shock.  If the luminosity ($L_j$) is low enough and/or the density of the stellar envelope is high enough, then the hydrodynamic jet is collimated and internal shocks might occur inside the progenitor  \citep{2013ApJ...777..162M,2011ApJ...740..100B, 2013PhRvL.111l1102M}.   In this model, inhomogeneities in the jet lead to internal shell collisions, higher  shells ($\Gamma_h$) catching slower  shells ($\Gamma_l$).  The kinetic energy of ejecta is partially dissipated via these internal shocks which take place at a distance of ${\small r_j=2\Gamma^2\,t_\nu}<R_\star$, where $t_v$ is the variability time scale of the central object, $\Gamma\simeq\sqrt{\Gamma_h\,\Gamma_l}$ is the bulk Lorentz factor of the propagating shock and $ R_\star$ is the radius of the progenitor's stellar surface.  The constraint $r_j<R_\star$ gives rise to those shocks inside the star.  The physical width of the internal shock is lower by a factor of $\Gamma$, i.e. $\Delta r_j = \Gamma\,t_v$.  These internal shocks are expected to be collisionless, so that particles may be accelerated.  In internal shocks the total energy density $U=1/(8\,\pi\,m_p)\,\Gamma^{-4}\,L_j\,t^{-2}_{\nu}$ is equipartitioned to generate and/or amplify the magnetic field $\epsilon_B=U_B/U=(B^2/8\pi)U$  \citep{2005AIPC..784..164P} and to accelerate particles $\epsilon_e=U_e/U$, where  m$_p$ is the proton mass. Then,  the magnetic field generated  at the shocks is written as
\be
B'= \epsilon_B^{1/2}\,\Gamma^{-2}\,L_j^{1/2}\,t^{-1}_\nu\,.
\label{mfield}
\ee
It is important to say that the strength of the magnetic field falls out of the shocked region  achieving  some Gauss and although its direction might be random, it is mostly transverse to the jet direction  \citep{raz10}.  From the causality condition,   the coherence length of such magnetic field is only the order of $\lambda_B\sim t_v$.  On the other hand, electrons are accelerated up to ultra-relativistic energies and then are cooled down rapidly in the presence of the magnetic field, producing the prompt emission by synchrotron radiation.   The opacity to Thomson scattering is $\tau_{th}'=\frac{\sigma_T}{4\pi\,m_p} \Gamma^{-3}\,L_j\,t^{-1}_\nu$ and  photons thermalize at a black body temperature with peak energy given by \citep{raz04b}
\be
T'_{\gamma}\simeq\frac{1.2}{\pi}\,\epsilon_e^{1/4}\, L_j^{1/4}\,\Gamma^{-1}\,t_v^{-1/2}\,,
\label{enph}
\ee
where $\sigma_T$ is the Thompson cross section.\\
Protons are also accelerated and cooled down in internal shocks  via electromagnetic (synchrotron radiation and inverse Compton (IC) scattering) and hadronic  (proton-photon and proton-proton interactions) channels.  Proton-photon and proton-proton interactions take place when accelerated protons interact  with  thermal keV  photons (eq. \ref{enph}) and proton density at the shock, ${\small n'_p=1/(8\,\pi\,m_p)\,\Gamma^{-4}\,L_j\,t^{-2}_{\nu}}$ \citep{PhysRevLett.87.171102}.  In both interactions HE charged pions and kaons are produced;  $p+\gamma/p \to X+\pi^{\pm}/K^{\pm}$, and subsequently neutrinos  $\pi^+\to \mu^++\nu_\mu\to e^++\nu_e+\bar{\nu}_\mu+\nu_\mu$ and $\pi^-\to \mu^-+\bar{\nu}_\mu\to e^-+\bar{\nu}_e+\nu_\mu+\bar{\nu}_\mu$.  In this approach, the neutrino created by these processes will lie in the TeV - PeV energy range \citep{2013PhRvL.111l1102M, 2014MNRAS.437.2187F, raz10}.\\
\section{Neutrino Effective Potential}
In this section we are going to compute the neutrino effective potential due to the magnetized and thermal shocked plasma, and the envelope of the star.
\subsection{Magnetized and thermal plasma}
\begin{figure*}
\centering
\includegraphics[width=0.4\textwidth]{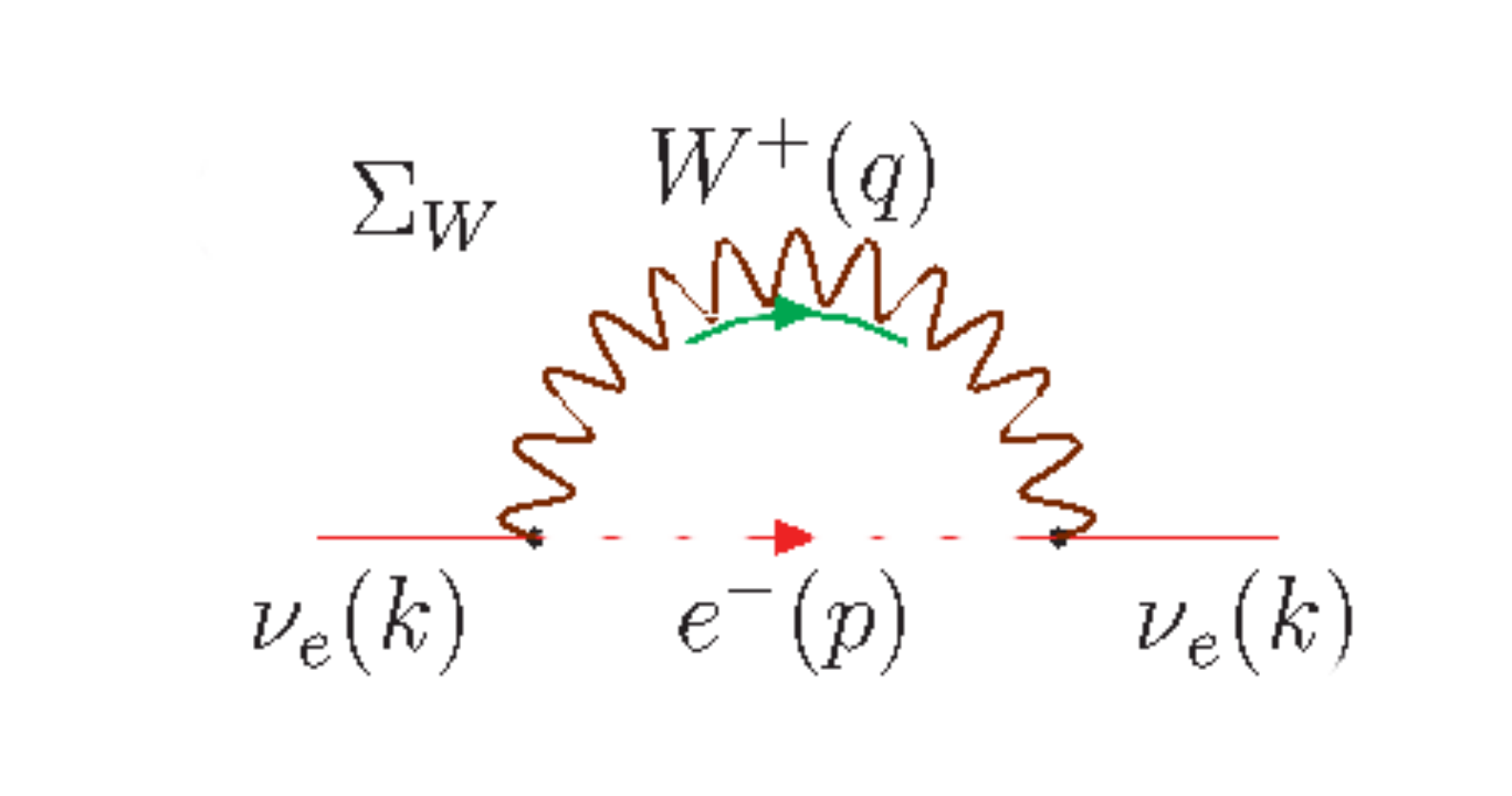}
\caption{W-exchange diagram of one-loop contribution to the neutrino self-energy in a magnetized medium. The dashed line represents the electron propagator $e^-(p)$, the solid line corresponds to the electron neutrino propagator $\nu_e(k)$ and the wiggly line is the W-boson propagator $W^+(q)$.   \label{diagrama}}
\end{figure*}

Recently, \citet{2014ApJ...787..140F} derived the neutrino self-energy and effective potential up to order $m_W^{-4}$ at strong, moderate and weak magnetic field approximation as a function of  temperature, chemical potential and neutrino energy for moving neutrinos along the magnetic field.  In this subsection, we will calculate the neutrino effective potential at the moderate and weak magnetic field limit for any direction of  neutrino propagation.  Therefore, following \citet{2014ApJ...787..140F} we will show the equations that are more relevant for deducing the neutrino effective potential.\\
The neutrino effective potential is calculated by means of the dispersion relation 
\be\label{diseq}
V_{eff}=k_0-|{\bf k}|\,,
\ee
where ${\bf k}$ is estimated through the neutrino field equation  in a medium \citep{1988NuPhB.307..924N, 1991NuPhB.349..754E}
\be
[ {\rlap /k} -{\Sigma}(k)]\psi_L=0\,,
\ee
 and  $\sigma_l(k)={\mathcal R} ( a_\parallel {\rlap /k}_\parallel + a_\perp {\rlap /k}_\perp + b {\rlap /u} + c {\rlap /b} ){\mathcal L}$ is obtained  from  the real part of its self-energy diagram. Here $k^\mu_\parallel=(k^0, k^3)$ and $k^\mu_\perp=(k^1,k^2)$ are the momentum along and perpendicular to the magnetic field, respectively,  $u^\mu$ stands for the 4-velocity of the center-of-mass of the medium given by $u^\mu = (1, {\bf 0})$, ${\mathcal R}=\frac12(1+\gamma_5)$ and $ {\mathcal L} = \frac12(1-\gamma_5)$ are the projection operators and a, b, and c are the Lorentz scalars which are functions of neutrino energy, momentum and magnetic field.   These scalars are calculated from the neutrino self-energy due to CC and NC interactions of neutrino with the background particles.  The effect of the magnetic field is introduced through the 4-vector $b^\mu$ which is given by $b^\mu = (0, {\hat {\bf b}})$ \citep{2014ApJ...787..140F}.   Using the Dirac algebra and  from the dispersion relation (eq. \ref{diseq}), the neutrino effective potential can be written as
\be
V_{eff}=b-c\,\cos\varphi-a_{\perp}|{\bf k}|\sin^2\varphi,
\label{poteff}
\ee
where $\varphi$ is the angle between the neutrino momentum and the magnetic field vector.   Otherwise, the effective potential that is applicable to the neutrino oscillations in matter is $V_{eff}=V_e-V_{\mu,\tau}$ which depends only on electron density \citep{wol78, 1992PhRvD..46.1172D}.  For that reason, although the one-loop neutrino self-energy  comes from three parts; the $W$-exchange, $Z$-exchange and tadpole \citep{2004PhRvD..70d3001B,1998PhRvD..58h5016E, 2009PhRvD..80c3009S,  2009JCAP...11..024S},  we will only consider the neutrino effective potential due to charged currents $\Sigma(k)=\Sigma_W(k)$.  We will use the finite temperature field theory formalism and the  Schwinger's propertime method to include the magnetic field \citep{1951PhRv...82..664S}.   From the W-exchange diagram (see fig. \ref{diagrama}), the  self-energy can be explicitly written as  
\be
-i\Sigma(k)={\mathcal R}\,\biggl[\frac{g^2}{2}\int\frac{d^4 p}{(2\pi)^4}\, \gamma_\mu S_{\ell}(p)\gamma_\nu\,
 \,W^{\mu \nu}(q)\biggr]\,{\mathcal L}\,,
\label{Wexch}
\ee
where $g^2=4\sqrt2 G_Fm_W^2$  is the weak coupling constant, $W^{\mu\nu}$ is the W-boson propagator that in unitary gauge can be written as \citep{1998PhRvD..58h5016E,  2009JCAP...11..024S}
\be\label{propW}
W^{\mu\nu}(q)=\frac{g^{\mu\nu}}{m^2_W}\biggl(1+\frac{q^2}{m^2_W}  \biggr)-\frac{q^\mu q^\nu}{m^4_W}+\frac{3ie}{2m^4_W}F^{\mu\nu}\,,
\ee
here m$_W$ is the W-boson mass, G$_F$ is the Fermi coupling constant,  $g^{\mu\nu}$ is the  metric tensor and $F^{\mu\nu}$ is the electromagnetic field tensor. From eq. (\ref{Wexch}), S$_l$(p) is the charged lepton propagator which is split in two propagators; one in presence of an uniform background magnetic field ($S^0_\ell(p)$) and the other in a magnetized medium ($S^\beta_\ell (p)$), then it can be written as 
\be
S_\ell (p) = S^0_\ell(p) + S^\beta_\ell (p)\,.
\label{slp}
\ee
We can express the charged lepton propagator  in presence of an uniform background magnetic field  as 
\be
i S^0_\ell(p) = \int_0^\infty e^{\Phi(p,s)} G(p,s)\,ds\,,
\label{sl0p}
\ee
where  the functions $\Phi(p,s)$ and $G(p,s)$ are written as
\bary
\Phi(p,s) &= &is(p_0^2 - m_\ell^2) - is[p^2_3 + \frac{\tan z}{z} p^2_\perp]\,,\cr
G(p,s)&=&\sec^2 z [{\rlap A/} + i {\rlap B/} \gamma_5 \cr
&&+m_\ell(\cos^2 z - i \sigma_l^3 \sin z \cos z)]\,,
\label{phaselu}
\eary
where $m_l$ is the mass of the charged lepton, $p^2_\parallel = p_0^2 - p_3^2$,  $p^2_\perp = p_1^2 + p_2^2$ are the projections  of the momentum on the magnetic field direction  and  $z= e{ B}s$, being $e$ the magnitude of the electron charge. Additionally,  the covariant  vectors are given as follows, $A_\mu = p_\mu -\sin^2 z (p\cdot u\,\, u_\mu - p\cdot b \,\,b_\mu)\,$, $B_\mu = \sin z\cos z (p\cdot u \,\,b_\mu - p\cdot b \,\,u_\mu)\,,$ and $\sigma_l^3 = \gamma_5 {\rlap /b} {\rlap /u}\,$.   The other term in eq. (\ref{slp})  (due to magnetized medium) is given by \citep{1996PhLB..383...87D}
\be
S^\beta_\ell(p) = i \eta_F(p\cdot u)\int_{-\infty}^\infty e^{\Phi(p,s)} G(p,s)
\,ds\,,
\label{slbp}
\ee
where $\eta_F(p\cdot u)$ contains the distribution functions of the particles in the medium which are given by:
\be
\eta_F (p \cdot u) = \frac{\theta(p\cdot u)}{e^{\beta(p\cdot u -
\mu_\ell)} + 1 } +
\frac{\theta(- p\cdot u)}{e^{-\beta(p\cdot u - \mu_\ell)} + 1}\,,
\label{eta}
\ee
where $\beta$ and $\mu_\ell$ are the inverse of the medium temperature and the chemical potential of the charged lepton.   By evaluating eq. (\ref{Wexch}) explicitly we obtain
\be
Re \Sigma(k)= {\mathcal R}\,[a_{\perp} \rlap /k_\perp + b \rlap /u + c \rlap /b]\, {\mathcal L}\,,
\ee
where the Lorentz scalars are given by \citep{2014ApJ...787..140F}
{\small
\bary
a_{\perp}=-\frac{\sqrt2G_F}{m_W^2}\biggl[ \biggl\{E_{\nu_e}(n_e-\bar{n}_e)+ k_3(n_e^0-\bar{n}_e^0)\biggr\}\hspace{2cm} \nonumber\\
\hspace{0.4cm}+\frac{eB}{2\pi^2}\int^\infty_0 dp_3\sum_{n=0}^\infty(2-\delta_{n,0}) \biggl (\frac{m_e^2}{E_n}- \frac{H}{E_n}\biggr)(f_{e,n}+\bar{f}_{e,n})\biggr],
\label{conaw}
\eary
}
{\small
\bary
b&=&\sqrt2 G_F \biggl[\biggl(1+\frac{E_{\nu_e}^2}{m_W^2}\biggr)(n_e-\bar{n}_e)+ \frac{ E_{\nu_e}k_3}{m_W^2}(n_e^0-\bar{n}_e^0)\nonumber \\
&&-\frac{eB}{\pi^2m_W^2} \int^\infty_0 dp_3 \sum_{n=0}^\infty(2-\delta_{n,0})E_{\nu_e}\biggl\{E_n\delta_{n,0} +\biggl(E_n \nonumber\\
&&\hspace{4.2cm}-\frac{m_e^2}{2E_n}\biggr)\biggr\}(f_{e,n}+\bar{f}_{e,n})\biggr]\nonumber\,, \\
\label{conbw}
\eary
}
and
{\small
\bary
c=\sqrt2 G_F\biggl[\biggl(1-\frac{k_3^2}{m_W^2}\biggr)(n_e^0-\bar{n}_e^0) -\frac{E^2_{\nu_e}}{m_W^2}(n_e-\bar{n}_e) \hspace{0.9cm} \nonumber\\
-\frac{eB}{\pi^2m_W^2} \int^\infty_0 dp_3\sum_{n=0}^\infty(2-\delta_{n,0}) E_{\nu_e}\biggl\{\biggl(E_n-\frac{m_e^2}{E_n}\biggr)\delta_{n,0}\nonumber \\
+\biggl(E_n-\frac32\frac{m_e^2}{E_n}-\frac{H}{E_n}\biggr)\biggr\}(f_{e,n}+\bar{f}_{e,n})\biggr].
\label{concw}
\eary
}
Here the electron number density  and electron distribution function are
\be
n_e(\mu, T, B)=\frac{eB}{2\pi^2}\sum_{n=0}^\infty (2 - \delta_{n,0}) \int_0^\infty  \frac{dp_3}{e^{\beta(E_{e,n}-\mu)} +1},
\label{den}
\ee
and
\be
f(E_{e,n},\mu)=\frac{1}{e^{\beta(E_{e,n}-\mu)} +1}\,,
\label{den1}
\ee
respectively, with $\bar{f}_{e,n}(\mu, T)=f_{e,n}(-\mu, T)$ and  $E_{e,n}=\sqrt{p_3^2+m_e^2+H}$ with $H=2neB$.  Solving the integral terms in eqs. (\ref{conaw}), (\ref{conbw}) and  (\ref{concw}) and replacing them in eq. (\ref{poteff}) we calculate the neutrino effective potential for two cases: the moderate and the weak magnetic field limit. 
\subsubsection{Moderate Magnetic field limit}
In the moderate field approximation ($ B/B_c \leq 1$), the Landau levels are discrete and can be described by sums  ($\sum_n$ with n=1, 2, 3 ..).  In this regime,  the neutrino effective potential  is written as
\begin{eqnarray}\label{Veffm}
V_{eff,is(m)}=\frac{\sqrt2\,G_F\,m_e^3 B}{\pi^2\,B_c}\biggr[\sum^{\infty}_{l=0}(-1)^l\sinh\alpha_l   \left[F_m-G_m\cos\varphi \right]\nonumber\\
-4\frac{m^2_e}{m^2_W}\,\frac{E_\nu}{m_e}\sum^\infty_{l=0}(-1)^l\cosh\alpha_l  \left[J_m-H_m\cos\varphi \right]  \biggr]
 \end{eqnarray}
where $\alpha_l=\beta\mu(l+1)$ and the functions F$_m$, G$_m$, J$_m$, H$_m$ are written in the appendix A. It is worth noting that as the magnetic field decreases the effective potential will depend less on the Landau levels.
\subsubsection{Weak Magnetic field limit}
In the weak field approximation ($ B/B_c \ll 1$),  all  levels are full and overlap each other.  In this regimen, sums over the Landau levels can be described and approximated  by an integral  $\sum_n\to\int {\small dn}$, then the effective potential does not depend on the Landau levels. The potential in this regimen can be written as 
\begin{eqnarray}\label{Veffw}
V_{eff,is(w)}=\frac{\sqrt2\,G_F\,m_e^3 B}{\pi^2\,B_c}\biggr[\sum^{\infty}_{l=0}(-1)^l\sinh\alpha_l  \left[F_w-G_w\cos\varphi \right]\nonumber\\
-4\frac{m^2_e}{m^2_W}\,\frac{E_\nu}{m_e}\sum^\infty_{l=0}(-1)^l\cosh\alpha_l \left[J_w-H_w\cos\varphi \right] 
\end{eqnarray}
where the functions F$_w$, G$_w$, J$_w$, H$_w$ are shown in the  appendix A.
\subsection{Density profiles of envelopes}
Models of density distributions in  CCSNe have been widely explored \citep{bet90, che89,woo93, shi90}.  We will use two models with density profiles $\rho \propto r^{-3}$ and $\rho \propto r^{-17/7}$. Explicitly, the  first model corresponds to a polytropic hydrogen envelope
\be
\hspace{0.7cm}\rho_1(r) = 4.0\times 10^{-6} \left( \frac{R_\star}{r} -1\right)^3 ~{\rm g~cm}^{-3}\,,
\label{dens-pro-A} \\
\ee
and the second model is a power-law fit with an effective polytropic index $n_{eff}=17/7$ as done for SN 1987A \citep{che89}
\bary
&&\rho_2(r) = 3.4\times 10^{-5} ~{\rm g~cm}^{-3}\cr
&&\times
\cases{ 
(R_\star/r)^{17/7}\,; \hspace{1cm}10^{10.8} ~{\rm cm}<r<r_b = 10^{12} ~{\rm cm}   &\nonumber \cr
(R_\star/r_b)^{17/7} (r-R_\star)^5/(r_b - R_\star)^5\,;\hspace{0.3cm} r>r_b\,.&\cr
} 
\label{dens-pro-B} \\
\eary
%.
In both cases, from the number density of electrons $N_e=N_a\,\rho(r)\, Y_e$,  the neutrino effective potential can be written as
\be\label{potencial_ss}
V_{eff,ss}=\sqrt2\,G_F N_e\,,
\ee
where N$_a=6.022\times 10^{23} g^{-1}$ is the Avogadro's number, $Y_e=$0.5 is the number of electrons associated per nucleon  and $\rho(r)$ is given by eqs. (\ref{dens-pro-A}) and (\ref{dens-pro-B}).
\section{Neutrino resonant oscillations}
When  neutrino oscillations take place in matter, a resonance could occur that would dramatically enhance the flavor mixing and could lead to a maximal conversion from one neutrino flavor to another.  This resonance depends on the effective potential and neutrino oscillation parameters.  The equation that determines the neutrino evolution in matter in the two and  three-flavor framework is \citep{2014MNRAS.442..239F}
\be
U\cdot \frac{1}{2E_\nu} \textbf{M}\cdot  U^\dagger + diag(V_{eff,k},\vec{0})\,,
\ee
where
\begin{equation}\label{p1}
\textbf{M}=
\cases{
(-\delta m^2, 0)			 			&  for two flavors\,,\cr
(-\delta m^2_{21}, 0,\delta m^2_{32})        &  for three flavors\,,\cr
}
\end{equation}
$\delta m^2_{ij}$ is the mass difference \citep{2007fnpa.book.....G}, $U$ is the  two- and  three-neutrino mixing matrix (see appendix B, eq. \ref{mixing}), $V_{eff,k}$ is the neutrino effective potentials calculated in section 3 (for $k$=is and ss) and $E_{\nu}$ is the neutrino energy.   We hereafter use the first and second line for two- and three-neutrino mixing, respectively, as written in  eq. (\ref{p1}).    From the conversion probabilities,  we obtain that the oscillation lengths are 
\begin{equation}\label{releng}
L_{res}= 4\pi E_\nu
\cases{
\frac{1}{\sqrt{(2 E_\nu V_{eff,k} -   \delta m^2 \cos 2\theta  )^2+ (\delta m^2 \sin 2\theta)^2}}, \cr
\frac{1}{\sqrt{(2 E_\nu V_{eff,k} -   \delta m_{32}^2 \cos 2\theta_{13}  )^2+ (\delta m_{32}^2 \sin 2\theta_{13})^2}},\cr
}
\end{equation}
with the resonance conditions
\begin{equation}\label{recond}
2 \times 10^6  E_\nu V_{eff,k} =
\cases{
\delta m^2 \cos 2\theta,\cr
\delta m_{32}^2 \cos 2\theta_{13}.\cr
}
\end{equation}
In addition to the resonance condition, the dynamics of the transition from one flavor to another must be  determined by adiabatic conversion through the adiabaticity  parameter \citep{2004mnpa.book.....M}
\begin{equation}\label{flip}
\gamma\equiv \frac{1}{2E_\nu\,\mid \frac{1}{V_{eff,k}}\,\frac {dV_{eff,k}}{dr}\mid_r}
\cases{
\delta m^2\,\sin2 \theta\,\tan2 \theta,\cr
\delta m_{32}^2\sin2 \theta_{13}\,\tan2 \theta_{13},\cr
}
\end{equation}
with  $\gamma\gg$ 1  or the flip probability  given by
\be
 P_f= e^{-\pi/2\,\gamma}\,.
 \label{flip_prob}
 \ee
By considering that the flux ratio of $\dot{N}_{\nu_\mu}\simeq\dot{N}_{\bar{\nu}_\mu}\simeq 2 \dot{N}_{\nu_e}\simeq2\dot{N}_{\bar{\nu}_e}$ is created in the internal shocks, neutrinos firstly oscillate in matter  due to the magnetized and thermal plasma and secondly oscillate to the star envelope.  In vacuum, after neutrinos have left the star, they start oscillating to the Earth.  Hence, from these three effects: internal shocks, envelope of the star and vacuum, the flavor ratio expected on Earth will be
\be
{\pmatrix
{
\nu_e   \cr
\nu_\mu   \cr
\nu_\tau   \cr
}_{Earth}}
=
{\pmatrix
{
P^*_{11}	  & P^*_{12}	  & P^*_{13}\cr
P^*_{21}	  & P^*_{22}	  & P^*_{23}\cr
P^*_{31}	  & P^*_{32}        & P^*_{33}\cr
}}
{\pmatrix
{
1  \cr
2  \cr
0   \cr
}_{c}}\,,
\label{matrixosc}
\ee
where the probabilities $P^*_{ij}$ are derived in appendix B.\\
The best fit values of the two neutrino mixing  are: \textbf{Solar Neutrinos}:  $\delta m^2=(5.6^{+1.9}_{-1.4})\times 10^{-5}\,{\rm eV^2}$ and $\tan^2\theta=0.427^{+0.033}_{-0.029}$\citep{aha11},  \textbf{Atmospheric Neutrinos}: $\delta m^2=(2.1^{+0.9}_{-0.4})\times 10^{-3}\,{\rm eV^2}$ and $\sin^22\theta=1.0^{+0.00}_{-0.07}$ \citep{abe11a} and   \textbf{Accelerator Neutrinos}: ${\small \delta m^2=0.5\,{\rm eV^2}}$ and  ${\small \sin^2\theta=0.0049}$ \citep{1994PrPNP..32..351Z, 1996PhRvL..77.3082A, 1998PhRvL..81.1774} . 
Combining solar, atmospheric, reactor and accelerator parameters, \textbf{the best fit values of the three neutrino mixing  are}, {\small ${\rm for}\,\,\sin^2_{13} < 0.053: \Delta m_{21}^2= (7.41^{+0.21}_{-0.19})\times 10^{-5}\,{\rm eV^2}$ and $\tan^2\theta_{12}=0.446^{+0.030}_{-0.029}$ and, {\rm for}\,\,$\sin^2_{13} < 0.04: \Delta m_{23}^2=(2.1^{+0.5}_{-0.2})\times 10^{-3}\,{\rm eV^2}$ and  $\sin^2\theta_{23}=0.50^{+0.083}_{-0.093}$}, \citep{aha11,wen10}.
\section{Results and Conclusions}
\begin{figure*}
\centering
\includegraphics[width=0.99\textwidth]{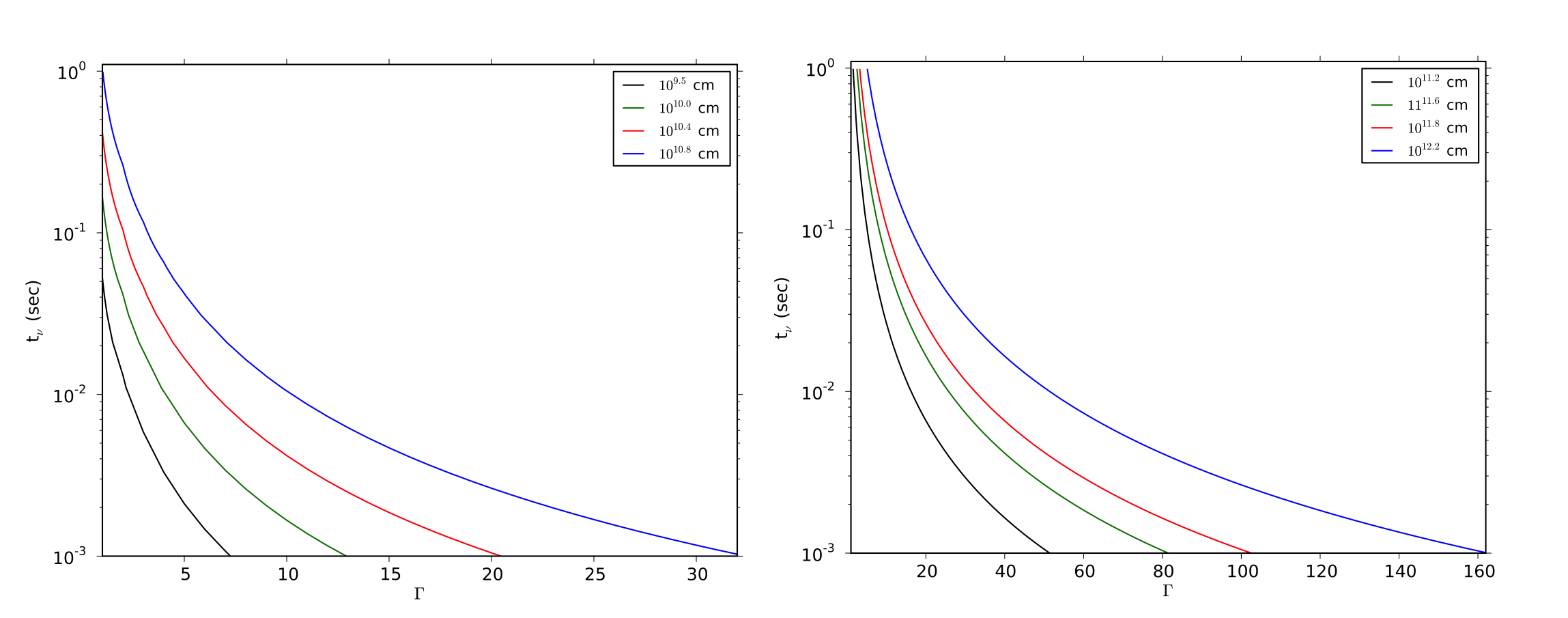}
\caption{Contour lines of variability time scale ($t_\nu$) and bulk Lorentz factor ($\Gamma$) as a function of the distance of the internal shocks ($r_j$) for which these shocks  take place inside the progenitors. We have considered progenitors such as  WR (left-hand figure) and BSG (right-hand figure) stars.\label{rj}}
\end{figure*}

\begin{figure*}
\centering
\includegraphics[width=0.99\textwidth]{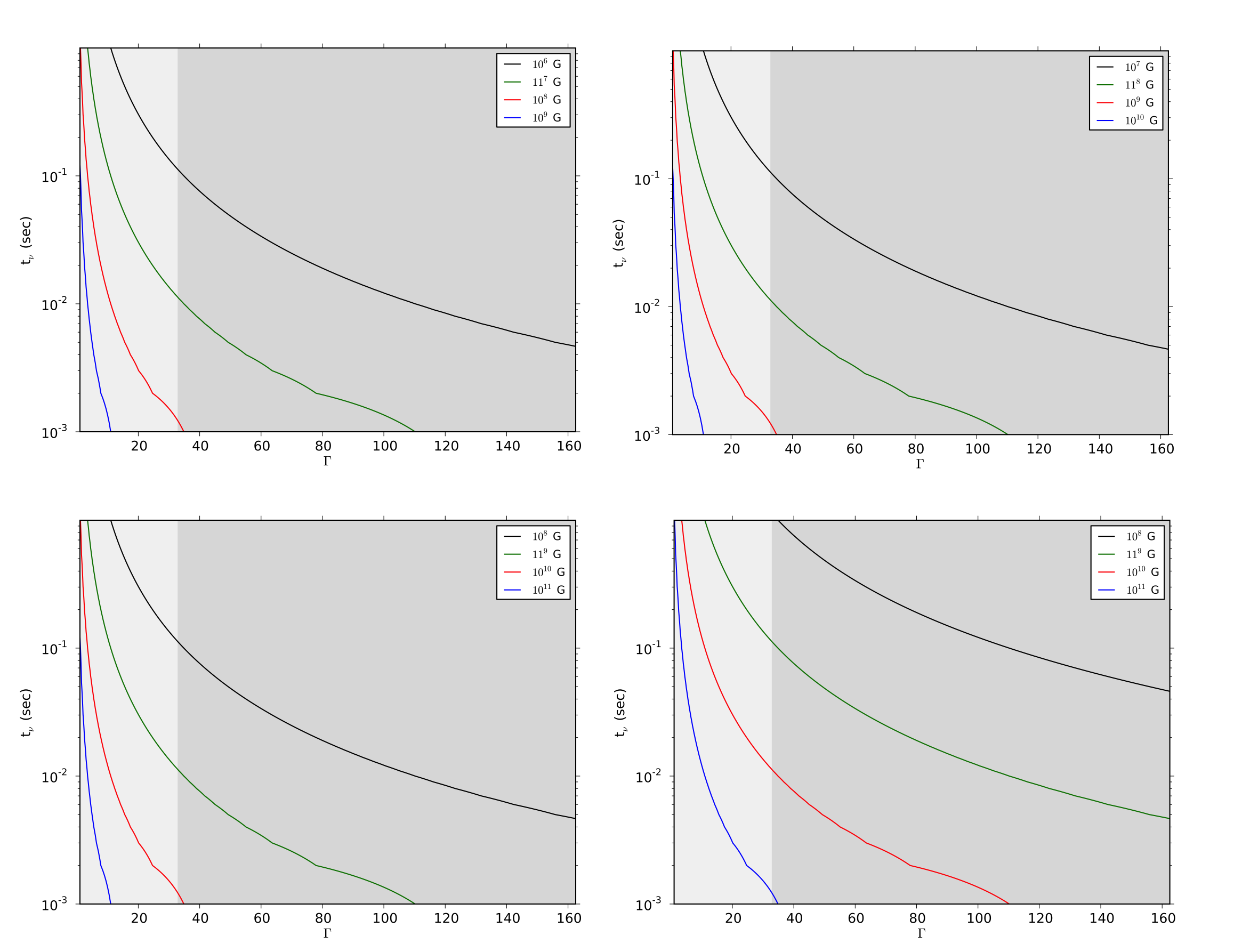}
\caption{Contour lines of variability time scale ($t_\nu$) and bulk Lorentz factor ($\Gamma$) as a function of magnetic field ($B'$) generated at internal shocks (eq. \ref{mfield}) inside a WR  (light gray background) and a BSG (dark gray background). In the left-hand figure above (below) a luminosity of 10$^{46}$ (10$^{50}$) erg/s was used, whereas in the right-hand figure above (below) it corresponds to a luminosity of 10$^{48}$ (10$^{52}$) erg/s.\label{Bj}}
\end{figure*}

\begin{figure*}
\centering
\includegraphics[width=0.99\textwidth]{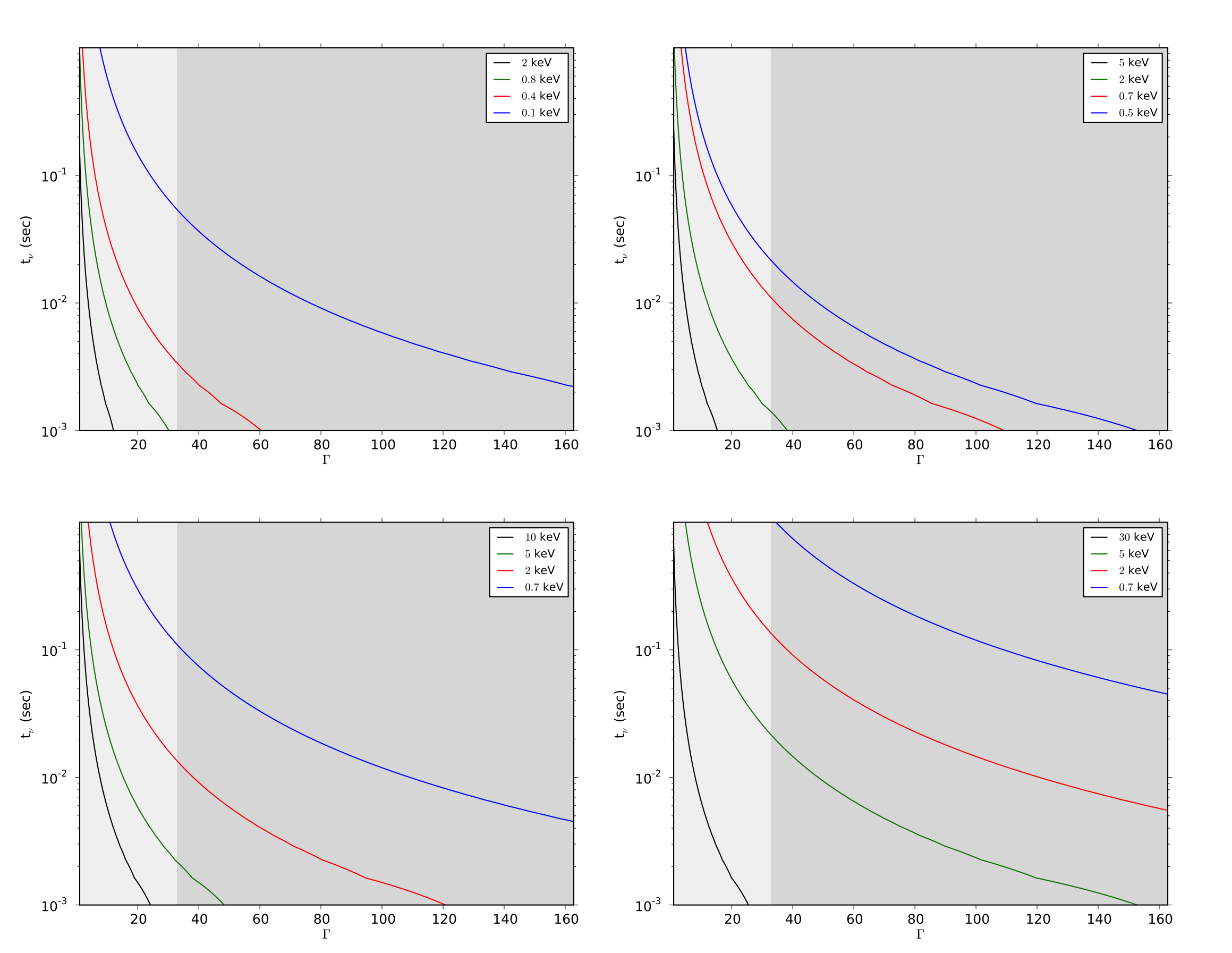}
\caption{Contour lines of variability time scale ($t_\nu$) and bulk Lorentz factor ($\Gamma$) as a function of synchrotron photons created in internal shocks and thermalized to a black body temperature ($T'_j$) (eq. \ref{enph}).   Once again we consider a WR  (light gray background) and a BSG (dark gray background). In the left-hand figure above (below) a luminosity of 10$^{46}$ (10$^{50}$) erg/s was used whereas in  the right-hand figure above (below) it corresponds to a luminosity of 10$^{48}$ (10$^{52}$) erg/s \label{Tj}}
\end{figure*}

\begin{figure*}
\centering
\includegraphics[width=\textwidth]{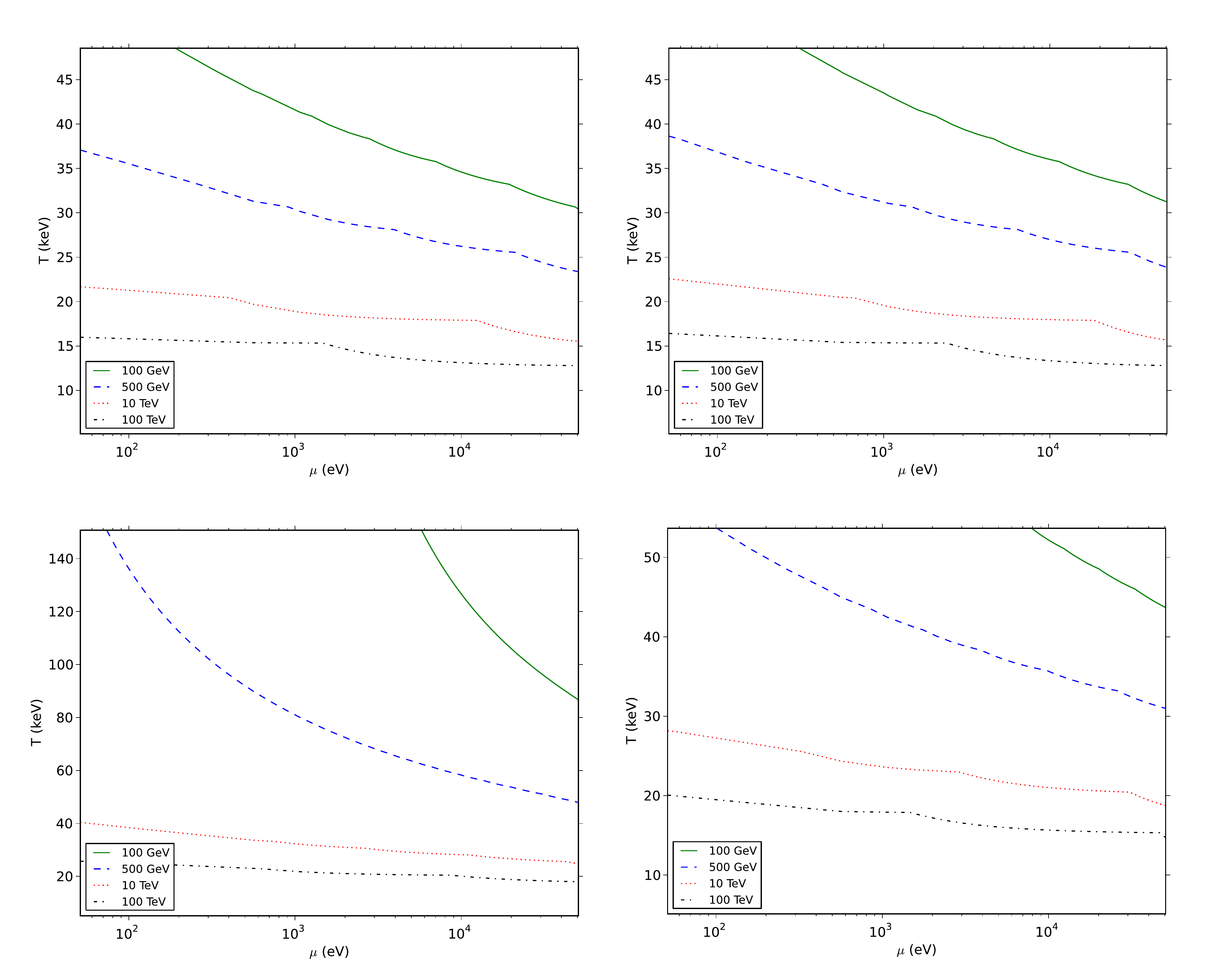}
\caption{Contour lines of temperature and chemical potential as a function of neutrino energy for which the resonance condition is satisfied.   We have used the neutrino effective potential at the moderate limit (eq. \ref{Veffm}) and the best-fit values of the two-neutrino mixing  (solar, top left; atmospheric, top right; and accelerator, bottom left) and three-neutrino mixing (bottom right).\label{res_M} }
\end{figure*}

\begin{figure*}
\centering
\includegraphics[width=\textwidth]{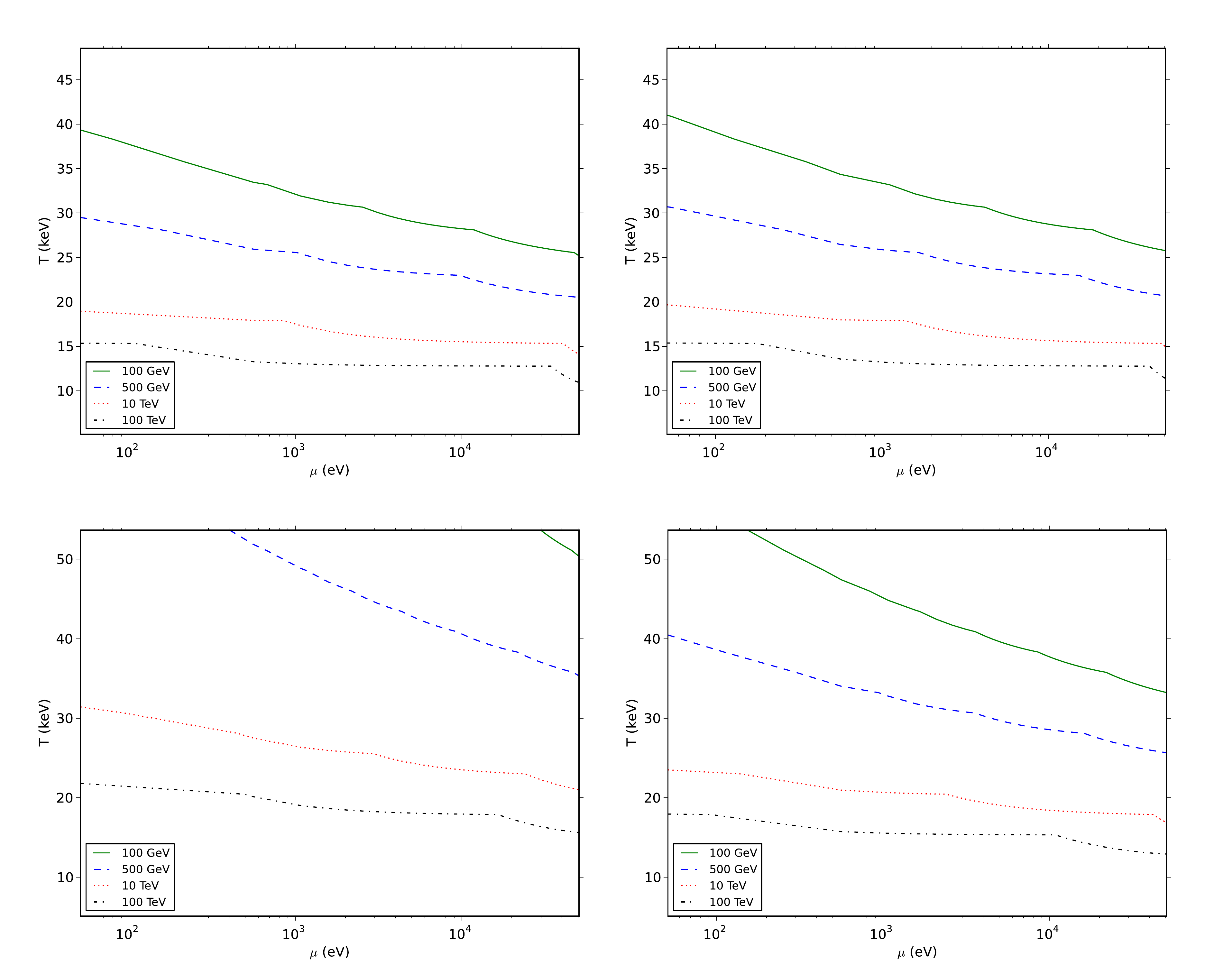}
\caption{Contour lines of temperature and chemical potential as a function of neutrino energy for which the resonance condition is satisfied.   We have used the neutrino effective potential at the weak limit (eq. \ref{Veffw}) and   the best-fit values of the two-neutrino mixing  (solar, top left; atmospheric, top right; and accelerator, bottom left) and three-neutrino mixing (bottom right).\label{res_W} }
\end{figure*}

\begin{figure*}
\centering
\includegraphics[width=0.5\textwidth]{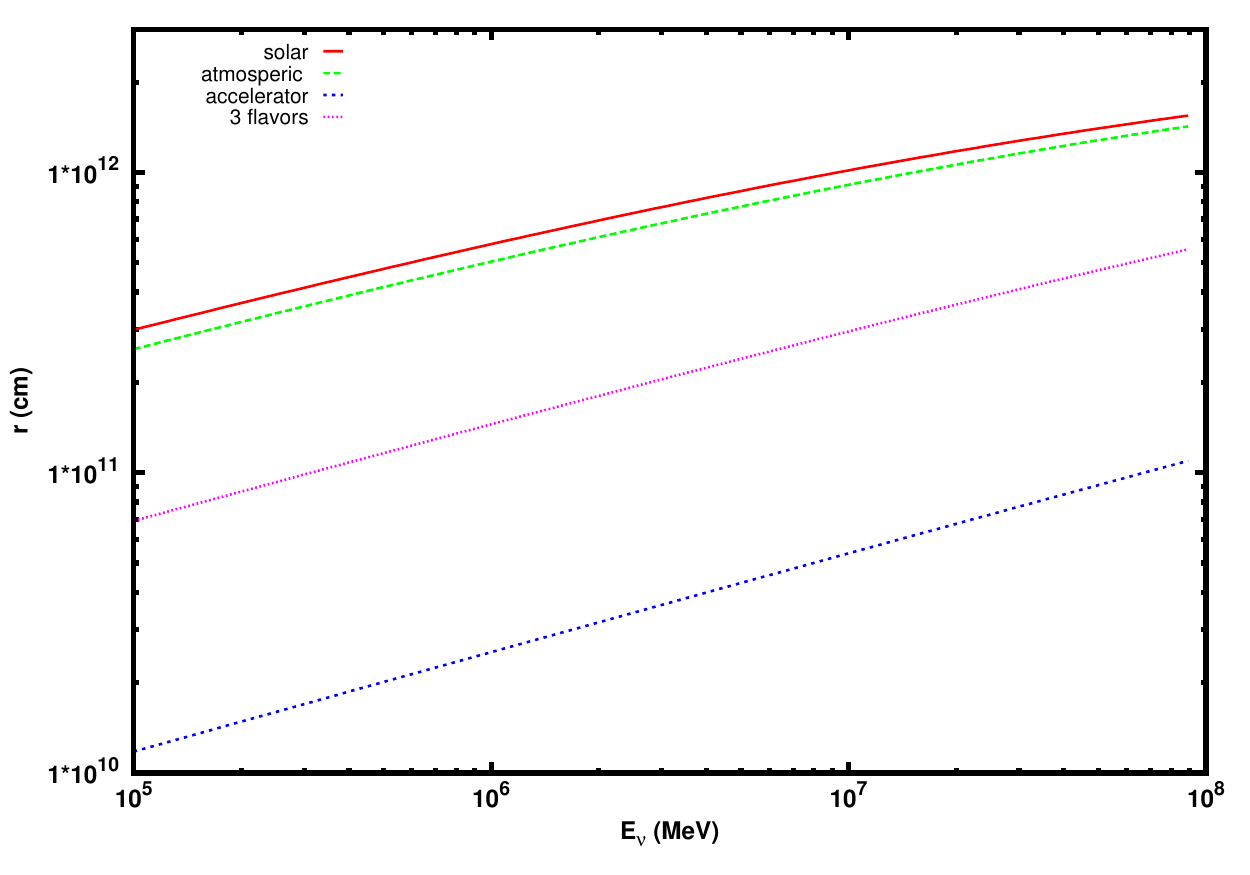}
\caption{Contour lines of distance and neutrino energy as a function of  neutrino oscillation parameters for which the resonance condition is satisfied.    We have used the neutrino effective potential generated by the envelope of star (eq. \ref{potencial_ss}) and  the best parameters of neutrino oscillation for  solar, atmospheric, accelerator and three flavors. \label{poten_ss} }
\end{figure*}

\begin{figure*}
\centering
\includegraphics[width=0.95\textwidth]{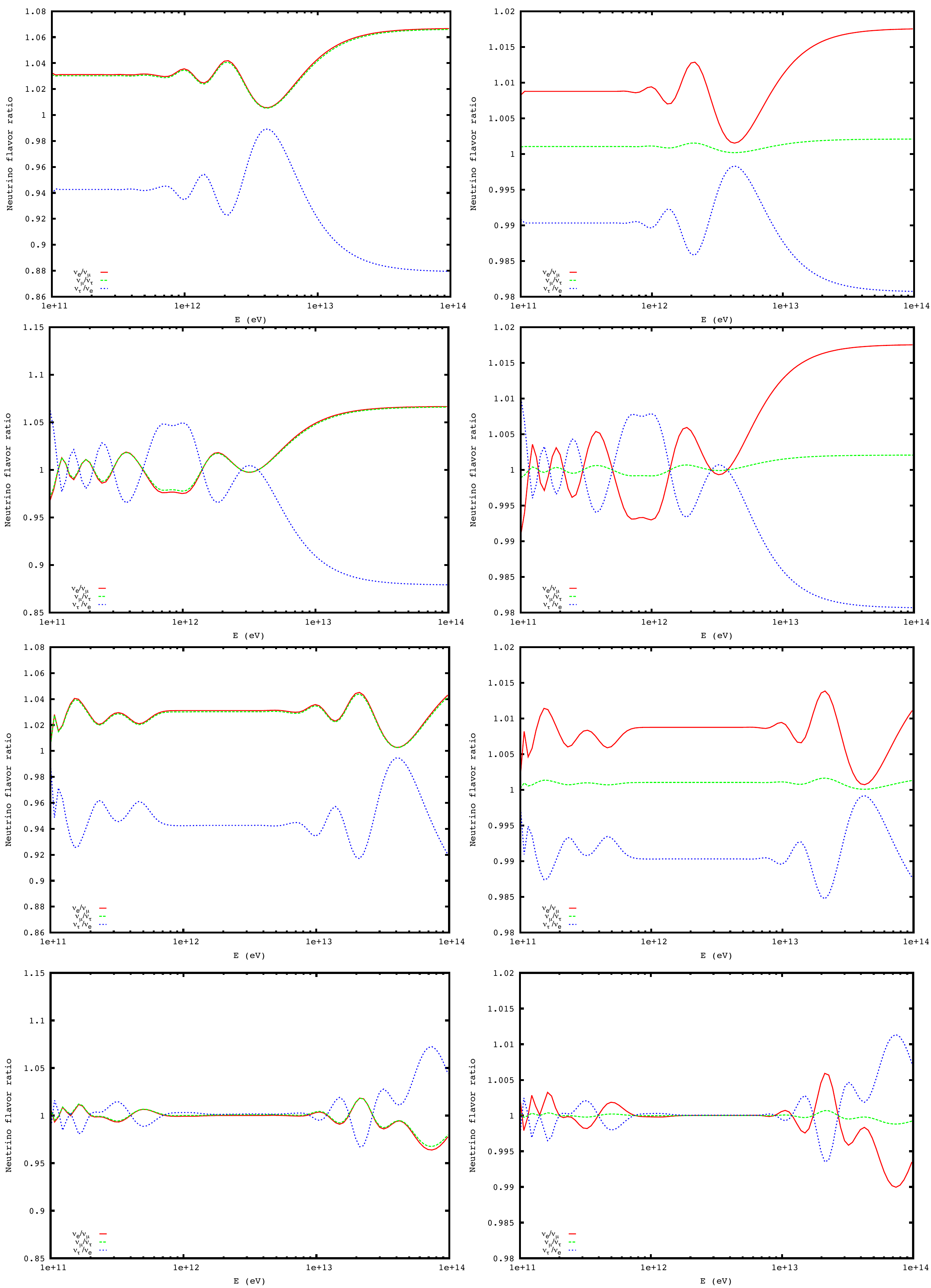}
\caption{Neutrino flavor ratio expected on Earth as a function of neutrino energy when these are created on the surface of a WR  (at $10^{11}$ and $10^{10.8}$ cm  for the upper and second panels, respectively) and a BSG (at $10^{12.3}$ and $10^{12}$ cm  for the third and bottom panels, respectively). We have used the neutrino effective potential at the moderate (second and bottom panels) and weak (upper and third panels)   field limit for $\theta_{13}$=11$^{\circ}$ (left) and $\theta_{13}=2^{\circ}$.  (right)\label{flavor_m}}
\end{figure*}

\begin{figure*}
\centering
\includegraphics[width=\textwidth]{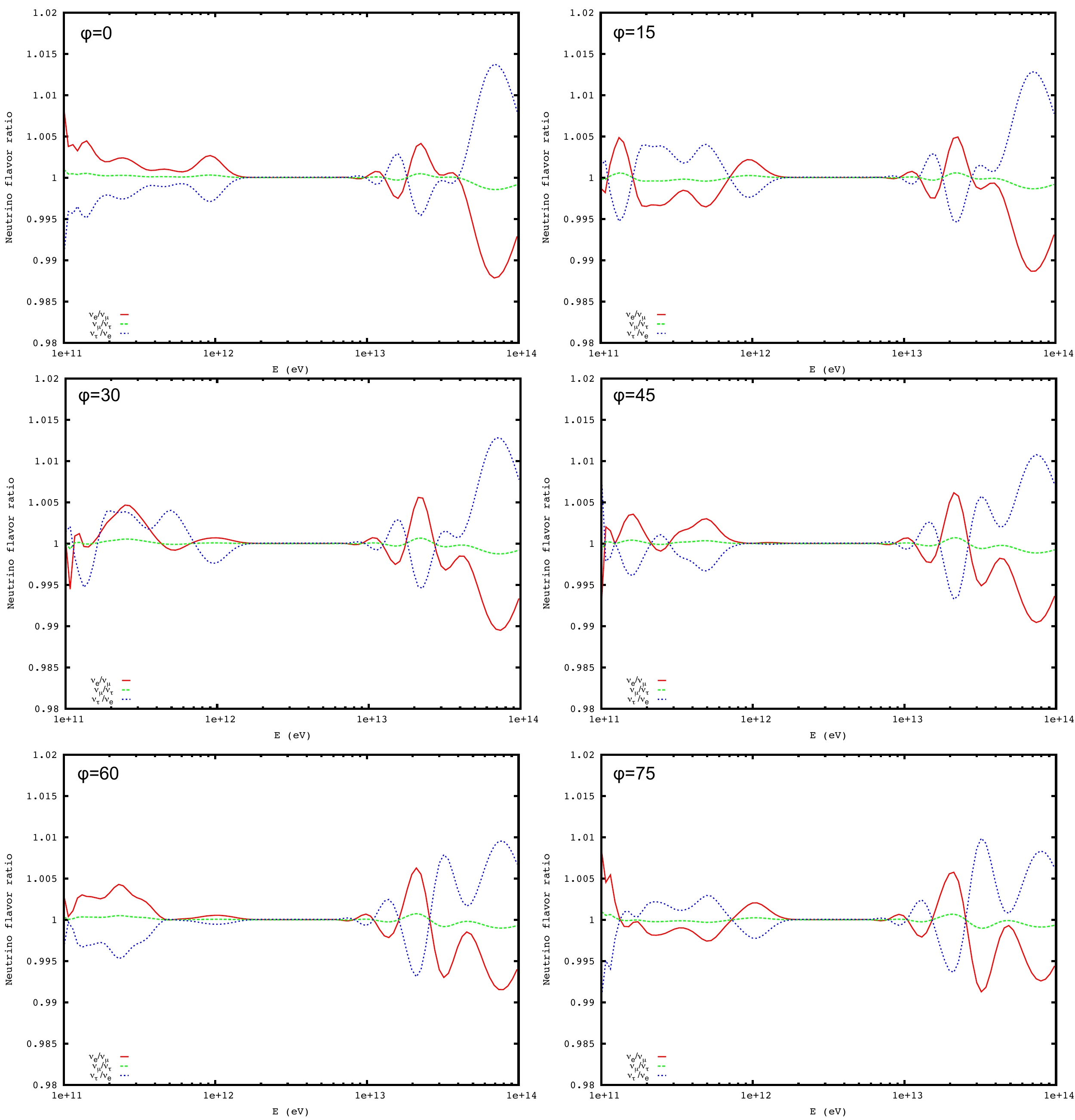}
\caption{Neutrino flavor ratio expected on Earth as a function of neutrino energy when these are created on the surface of a BSG. We have considered that internal shocks take place at $r_j=10^{12}$ cm with a physical width $\Delta r_j=2\times 10^{11}$ cm and the magnetic field is oriented to different angles  $0^\circ\leq \varphi \leq 75^\circ$ concerning to neutrino direction.  We have used the neutrino effective potential at the moderate-field limit (eq. \ref{Veffm}) and $\theta_{13}=2^{\circ}$. \label{flavor_am}}
\end{figure*}

\begin{figure*}
\centering
\includegraphics[width=\textwidth]{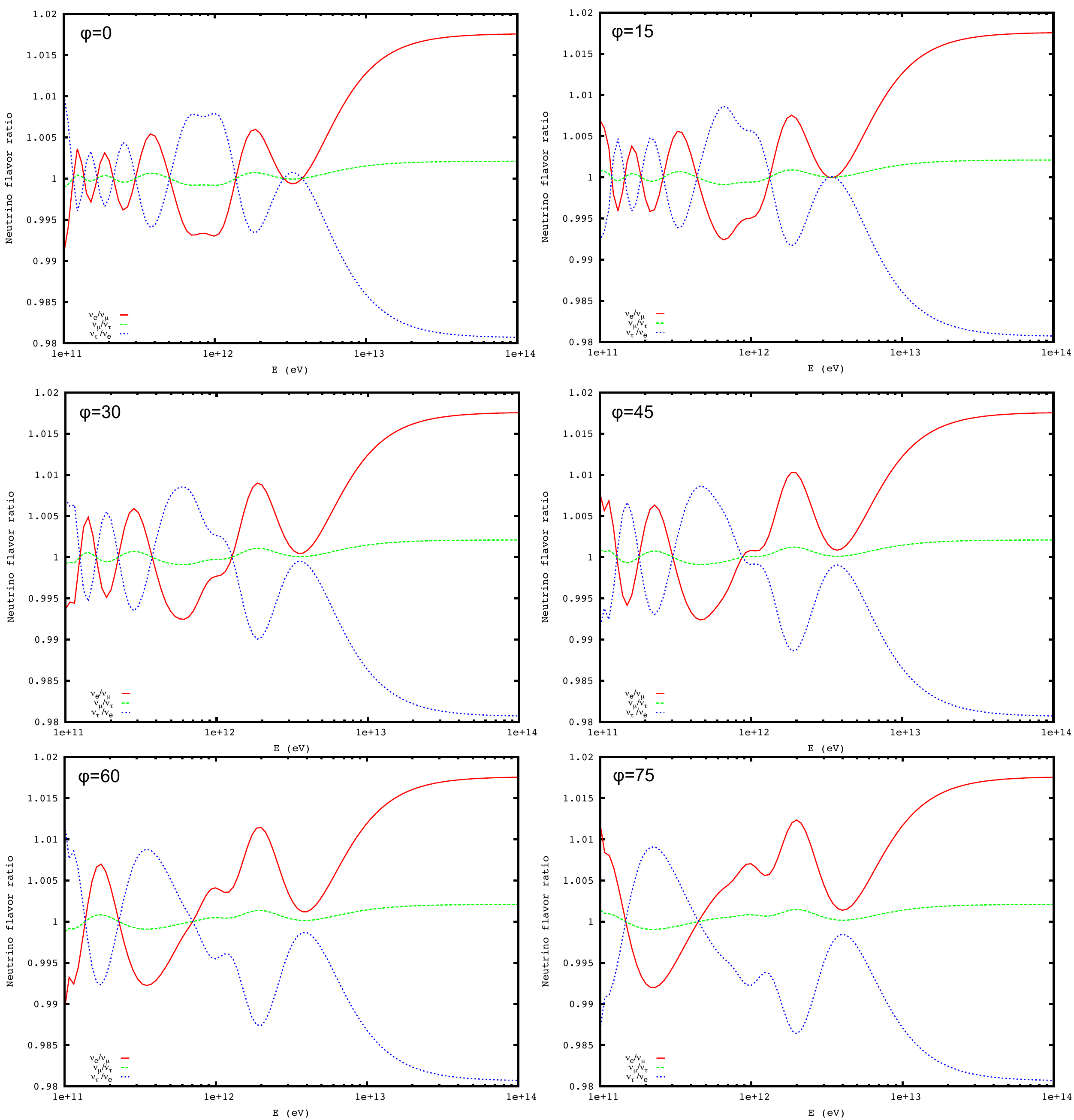}
\caption{Neutrino flavor ratio expected on Earth as a function of neutrino energy when these are created on the surface of a BSG. We have considered that internal shocks take place at $r_j=10^{12}$ cm with a physical width $\Delta r_j=2\times 10^{11}$ cm and the magnetic field is oriented to different angles  $0^\circ\leq \varphi \leq 75^\circ$ concerning neutrino direction.  We have used the neutrino effective potential at the moderate-field limit (eq. \ref{Veffw}) and $\theta_{13}=2^{\circ}$. \label{flavor_aw}}
\end{figure*}

In this analysis we have considered  HE neutrinos created in the energy range of 100 GeV$\leq  E_\nu \leq$ 100 TeV \citep{2013PhRvL.111l1102M, 2014MNRAS.437.2187F,raz10} and also we have assumed (in the CCSNe-GRB connection) progenitors such as Wolf-Rayet (WR) and  blue supergiant (BSG) stars with radii $R_\star=10^{11}$ cm and $R_\star=3 \times 10^{12}$ cm, respectively, with formation of jets leading to internal shocks  inside of them.\\
In internal shocks, energy is equipartitioned to generate and/or amplify the magnetic field and to accelerate particles.   Electrons and protons are expected to be accelerated in these shocks,  and after to be cooled down by synchrotron radiation, inverse Compton and hadronic processes (p$\gamma$ and p-hadron interactions).  Photons produced by electron synchrotron radiation are  thermalized at keV energies and serve as targets for production of HE neutrinos through  K$^\pm$, $\pi^\pm$ and $\mu^\pm$ decay products in the proton-$\gamma$ and proton-hadrons interactions.  Therefore, this plasma is endowed with a magnetic field and made of protons, mesons, electrons, positrons, photons and neutrinos.\\
First of all, we consider those internal shocks that take place inside progenitors (r$_j< R_\star$), as plotted in fig. \ref{rj}.  In this figure, we show the contour lines of bulk Lorentz factors and variability time scales for different internal shock radii.  In these plots we observe that for a typical value of  variability in the range of $10^{-3}\,{\rm s} \leq t_\nu\leq 1\,{\rm s}$, the values of bulk Lorentz factors are $\Gamma \leq 34$ for a WR (left-hand figure) and $\Gamma\leq 163$ for a BSG (right-hand figure). Taking into account internal shocks at  $r_j=10^{10.8}$ cm (left-hand figure) and  $r_j=10^{12.2}$ cm (right-hand figure) we can see that the physical width of the internal shocks is restricted to $\Delta r_j \leq 1.8\times 10^{10}$ cm and $\Delta r_j \leq 2.5\times 10^{11}$ cm, respectively.  Once obtained the values of $\Gamma$ and t$_\nu$ for internal shocks inside the progenitor we compute the range of values associated to the magnetic field and temperature of the plasma, as shown in figs. \ref{Bj} and \ref{Tj}, respectively. We plot the contour lines of the magnetic fields (fig. \ref{Bj}) and thermalized  photons  (fig. \ref{Tj}) for values of  luminosity in the range $10^{46}\, {\rm erg/s} \leq L_\gamma \leq10^{52}\, {\rm erg/s}$.  Colors in light- and dark-gray backgrounds represent the regions of  a WR and a BSG, respectively.\\
In figs. \ref{Bj} and \ref{Tj} one can see that the values of magnetic field and thermalized photons lie in the ranges $10^6\, \rm{G} \leq B \leq 10^{11}\, \rm{G}$ and $0.1\, \rm{keV} \leq T \leq 30\, \rm{keV}$, respectively. It is important to clarify that the magnetic field amplified in the internal shocks falls out of them to the magnetic field endowed by the progenitor (black hole (BH) or magnetar) \citep{raz10}.\\ 
Following \citet{2014ApJ...787..140F} and taking into account that the range of neutrino energy considered is larger than the W-boson mass (E$^2_\nu \geq m^2_W$),  we have obtained the neutrino effective potential up to an order $m^{-4}_W$ in the moderate below $B/B_c \sim 10^{-5}$ and weak  $B/B_c \sim 10^{-13}$ regime as a function of the observable quantities in the internal shocks: thermalized photons,  magnetic field, neutrino energy and angle between the direction of neutrino propagation and the magnetic field.  We plot the neutrino effective potential in both limits (moderate and weak field limits)  as shown in fig. \ref{potential}. 
The neutrino effective potential at moderate limit (figures above) and weak limit (figures below) are plotted for a magnetic field in the range of $10^{-6}\, B_c< B < 10^{-4}\, B_c$ and $10^{-13}\, B_c< B < 10^{-12}\, B_c$, respectively. In both cases,  we use the values of temperature T=  (20, 24, 27 and 30) keV,  angle $\varphi$=  ($0^\circ$, $30^\circ$, $60^\circ$ and $90^\circ$) and the neutrino energy E$_\nu$=10 TeV.   The neutrino effective potential at the weak limit is smaller than at the moderate limit.  It is worth mentioning that in the range of the magnetic field considered, the contribution of Landau levels  to the effective potential at the moderate-field limit is not significant  due to $\sum^\infty_{n=1}\lambda_n\,K_i(\sigma_l\lambda_n)\sim 0$ for $\lambda_n=\sqrt{1+2\,n\,B/B_c}$ and  $\sigma_l=\beta m_e(l+1)$.  From these plots one can observe that the neutrino effective potential is positive, therefore  due to its positivity  ($V_{eff,is(k)}>$ 0) for k= m and w,  neutrinos can oscillate resonantly.   From the resonance condition (eq. \ref{recond}) and the neutrino effective potential at moderate (eq. \ref{Veffm}) and weak (eq. \ref{Veffw})  limit,  we plot the contour lines of temperature and chemical potential as a function of neutrino energy for which the resonance condition is satisfied, as shown in figs. \ref{res_M} and \ref{res_W}, respectively. From these figures, one can see that temperature is a decreasing function of chemical potential and neutrino energy.  As neutrino energy increases,  temperature decreases steadily.  Considering the values of neutrino energy ($E_\nu$ =100 GeV, 500 GeV, 10 TeV and 100 TeV) and  $\varphi=90^\circ$, we see that  the temperature and chemical potential are in the range 10 keV to $\sim$100 keV and 60 eV to 50 keV, respectively. For instance,  taking into account a neutrino energy of 10 TeV,  from fig. \ref{res_M} we can see that temperature lies in the range  22.2 to 15.3 keV for solar, 23.4 to 16.2 keV for atmospheric, 40 to 26.4 keV for accelerator and  28.3 to 15.4 keV for three-neutrino parameters, and  as shown in fig. \ref{res_W}, temperature lies in the range 18.3 to 14.1 keV for solar, 19.8 to 15.1 keV for atmospheric, 32.1 to 21.9 keV for accelerator and  24.5 to 18.3 keV for three-neutrino parameters.   In addition, we have obtained the resonance lengths which are shown in table \ref{res_len}.  As shown in this table, the resonance lengths lie in the range $l_{res}\sim$ (10$^{10}$ to $10^{13}$) cm, hence depending on the progenitor associated and the oscillation parameters, neutrinos would leave the internal shock region in different flavors of 1:2:0. For instance, taking into account the parameters of three-neutrino mixing, neutrinos with energy less than 0.5 (10) TeV will oscillate resonantly with a resonance length equal or less than the radius of the progenitor, either a WR or BSG.   Considering parameters of accelerator experiments, neutrino energy around 100 TeV  will oscillate  in a BSG star before leaving it. \\
\begin{table}
\begin{center}\renewcommand{\tabcolsep}{0.18cm}
\renewcommand{\arraystretch}{0.89}
\begin{tabular}{ccrlc} \hline
Energy                       &                                                      & $l_{res}$& (cm)& \\
{\small (TeV)}             &  {\small Solar}                               &   {\small Atmosph.}                         &    {\small Accelerat.}                   & {\small Three flav.}                  \\\hline
{\small $10^{-2}$}          &   {\small  4.8 $\times 10^{10}$}     &   {\small 1.2 $\times 10^{9}$}         &    {\small 7.1 $\times 10^7$}       & {\small 2.6 $\times 10^9$}       \\
{\small 0.5 }               &    {\small 2.4 $\times 10^{12}$}     &   {\small 5.9 $\times 10^{10}$}       &    {\small 3.6 $\times 10^9$}       & {\small 1.3 $\times 10^{11}$}   \\
{\small 10}                 &    {\small 4.8 $\times 10^{13}$}     &   {\small 1.2 $\times 10^{12}$}       &    {\small 7.1 $\times 10^{10}$}   &  {\small 2.6 $\times 10^{12}$}    \\
{\small $10^2$}             &    {\small 4.8 $\times 10^{14}$}     &   {\small 1.2 $\times 10^{13}$}       &    {\small  7.1 $\times 10^{11}$}   & {\small 2.6 $\times 10^{13}$}  \\\hline
\end{tabular}
\end{center}
\caption{\small\sf Resonance lengths  of HE neutrinos for the  best fit parameters of the two- and three-neutrino mixing.}
\label{res_len}
\end{table}
As the dynamics of resonant transitions is not only determined by the resonance condition, but also by adiabatic conversion, we analyze the flip probability (eq. \ref{flip_prob}) to find the regions for which neutrinos can oscillate resonantly.  First of all we derive the neutrino effective potential as the function of magnetic field $dV_{eff}/dr=\partial V_{eff}/\partial B\times \partial B/\partial r$, and assume that at internal shocks ($10^{10.8}$ cm for WR and $10^{12}$ cm for BSGs),  magnetic fields change a 10\% of any variation around  the radius shock scale. For instance, for a WR star,   $\partial B/\partial r=0.1\times10^{-4} B_c/10^{10.8} {\rm cm}=  6.99\times 10^{-3}$ Gauss/cm.   We plot the flip probability as a  function of neutrino energy for two  and three flavors (fig \ref{flip}). We divide each plot of flip probability in three regions in order to analyze the whole range of probabilities: less than 0.2 (P$_\gamma \leq$ 0.2, a pure adiabatic conversion),  between 0.2 and 0.8 (0.2 $<$ P$_\gamma$ $<$ 0.8 represents the transition region) and greater than 0.8 (P$_\gamma \geq$ 0.8 is a strong violation of adiabaticity)\citep{dig00}. In fig. \ref{flip}, we use two flavors: solar  (left-hand figure above), atmospheric (right-hand figure above), accelerator  (left-hand figure below) and three flavor (right-hand figure below).    When we use solar parameters,  a pure adiabatic conversion occurs in a WR (BSG) star for neutrino energies of less than $10^{10.5}$ (10$^{11.7}$) eV and $10^{11.5}$ (10$^{12.75}$) eV which are endowed with $B=10^{-4} B_c $ and $B=10^{-13} B_c$, respectively.  Considering atmospheric parameters,  only a pure adiabatic conversion takes place in a WR (BSG) star for neutrino energies less than $10^{13.6}$ (10$^{14.8}$) eV and $10^{14.5}$ ($>$ 10$^{15}$) eV which are endowed with $B=10^{-4} B_c$ and $B=10^{-13} B_c$, respectively.  Taking into account  accelerator parameters,  a pure adiabatic conversion happens in a WR (BSG) star for neutrino energies of less than $10^{11.8}$ (10$^{13.1}$) eV and $10^{12.7}$ $>$ 10$^{15}$) eV with $B=10^{-4} B_c $ and $B=10^{-13} B_c $, respectively and  once again considering three neutrino mixing,  a pure adiabatic conversion occurs  in a WR (BSG) star for neutrino energy of less than $10^{11.1}$ (10$^{12.5}$) eV and $10^{12.1}$ (10$^{3.5}$) eV with $B=10^{-4} B_c $ and $B=10^{-13} B_c $, respectively. Higher energies to those considered are found in regions of transition and/or those prohibited. \\
In addition,  we have studied the HE neutrino oscillations from the neutrino effective potential generated in the star envelope (eq. \ref{potencial_ss}), as shown in fig. \ref{poten_ss}.  From the resonance condition (eq. \ref{recond}), we obtain the contour plots of radius as a function of neutrino energy. One can see that for neutrino energy in the range 100 GeV$<E_\nu<$ 100 TeV the radius lies  in the range   $10^{10.8}$ cm $<\,r<\, 10^{12.5}$ cm. The flip probability for neutrino oscillations in the envelope of a star was studied by \citet{2014MNRAS.437.2187F}.  The author has plotted this probability as a function of neutrino energy for  density profiles [A] (eq. \ref{dens-pro-A}) and  [B] (eq. \ref{dens-pro-B}) and neutrino oscillation parameters. From this analysis,  \citet{2014MNRAS.437.2187F} showed that neutrinos can oscillate depending on their energy and the parameters of neutrino experiments, obtaining that neutrino with energies above  dozens  of TeV can hardly oscillate.\\
Finally, considering a flux ratio $\dot{N}_{\nu_\mu}\simeq\dot{N}_{\bar{\nu}_\mu}\simeq 2 \dot{N}_{\nu_e}\simeq2\dot{N}_{\bar{\nu}_e}$, we estimate the neutrino flavor ratio coming from the surface of a WR  and BSG to Earth, as shown in fig. \ref{flavor_m}.   In this estimation, we take into account the contribution of thermal and magnetized plasma  at moderate and weak $B$ limit generated by internal shocks; at  $10^{10.8}$ cm (second panel), $10^{11}$ cm (upper panel), $10^{12}$ cm (bottom panel) and $10^{12.3}$ cm (third panel),  the effective potential due to the envelope of star and oscillation neutrinos in vacuum, due to the path up to Earth.  In this figure we take into account two  values of $\theta_{13}$ mixing angle, 2$^\circ$ (left column) and 11$^\circ$ (right column).  As shown,  one can observe that a nonsignificant deviation of the standard ratio ($\phi_{\nu_e}/\phi_{\nu_\mu}$:$\phi_{\nu_\mu}/\phi_{\nu_\tau}$:$\phi_{\nu_\tau}/\phi_{\nu_e}$ ; 1:1:1) is expected, less than 10 \% for $\theta_{13}=11^\circ$ and  only 2 \% for $\theta_{13}=2^\circ$.  In addition, we plot the neutrino flavor ratio expected on Earth as a function of neutrino energy  when the magnetic field is oriented to different angles  $0^\circ\leq \varphi \leq 75^\circ$ concerning neutrino direction, as shown in figs.  \ref{flavor_am} and  \ref{flavor_aw}. In fig. \ref{flavor_am} we consider  the neutrino effective potential at the moderate-field limit and internal shocks at $r_j=10^{12}$ cm with a physical width $\Delta r_j=2\times 10^{11}$ cm and in fig. \ref{flavor_aw}, we consider the neutrino effective potential at the weak-field limit and internal shocks at $r_j=10^{10.8}$ cm with a physical width $\Delta r_j=1.5\times 10^{10}$ cm.   From both plots, we can see that although the neutrino flavor ratio changes at different angles, distances of internal shocks, strength of magnetic field (moderate and weak limit) and neutrino energy in the range $10^{11}\, {\rm eV} \leq E_\nu\leq 10^{14}\, {\rm eV}$, this  flavor ratio expected on Earth  lies between 0.98 and 1.02, hence we can conclude that the directionality of magnetic fields does not affect our results.  Although currently neutrino oscillations can hardly be detected, new techniques in the near future will allow us to perceive these oscillations and put limits on the neutrino mixing angles. Finally, it is worth noting that the estimated values of the bulk Lorentz factor, in particular those relying on variability time measurements, are only raw approximations, and variations by a factor of a few cannot be ruled out by existing data.\\
\section*{Acknowledgements}
We are thankful to the anonymous referee for a critical reading of the paper and valuable suggestions that helped improve the quality and clarity of this work.  We also thank  A. M. Sodelberg, J. Nieves, B. Zhang,  K. Murase, W. H. Lee, F. de Colle, E. Moreno and A. Marinelli for useful discussions.   NF gratefully acknowledges a Luc Binette-Fundaci\'on UNAM postdoctoral fellowship. This work was supported by the projects IG100414 and Conacyt 101958.
% 
%\bibliographystyle{mn2e}
%\bibliography{Bib_osc}
%

%
\begin{figure*}
\centering
\includegraphics[width=\textwidth]{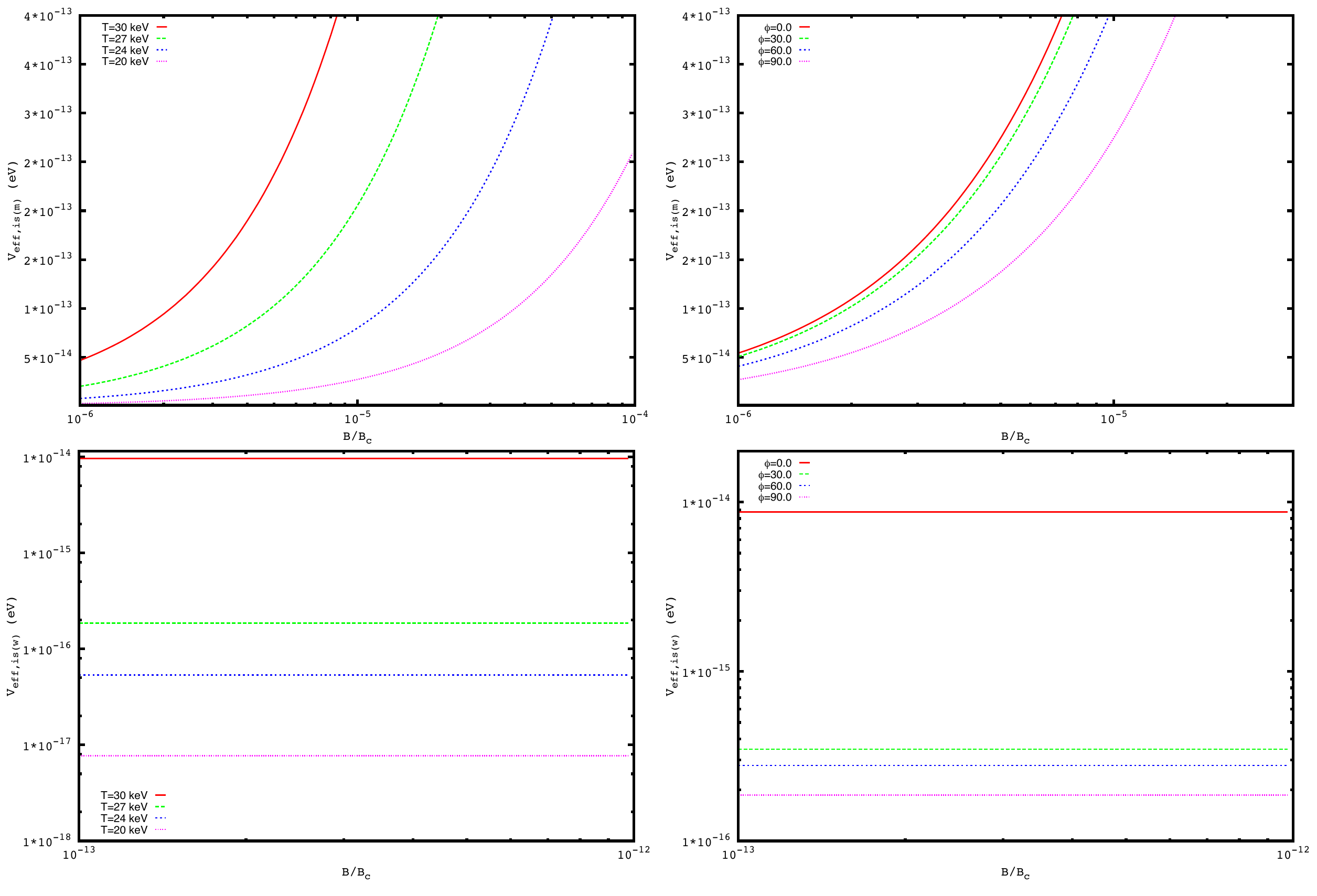}
\caption{Neutrino effective potentials at moderate (top) and weak (bottom) limits are plotted as a function of magnetic field for temperatures at keV energies (left) and different angles between the direction of neutrino propagation and magnetic field (right).  \label{potential}}
\end{figure*}
\begin{figure*}
\centering
\includegraphics[width=\textwidth]{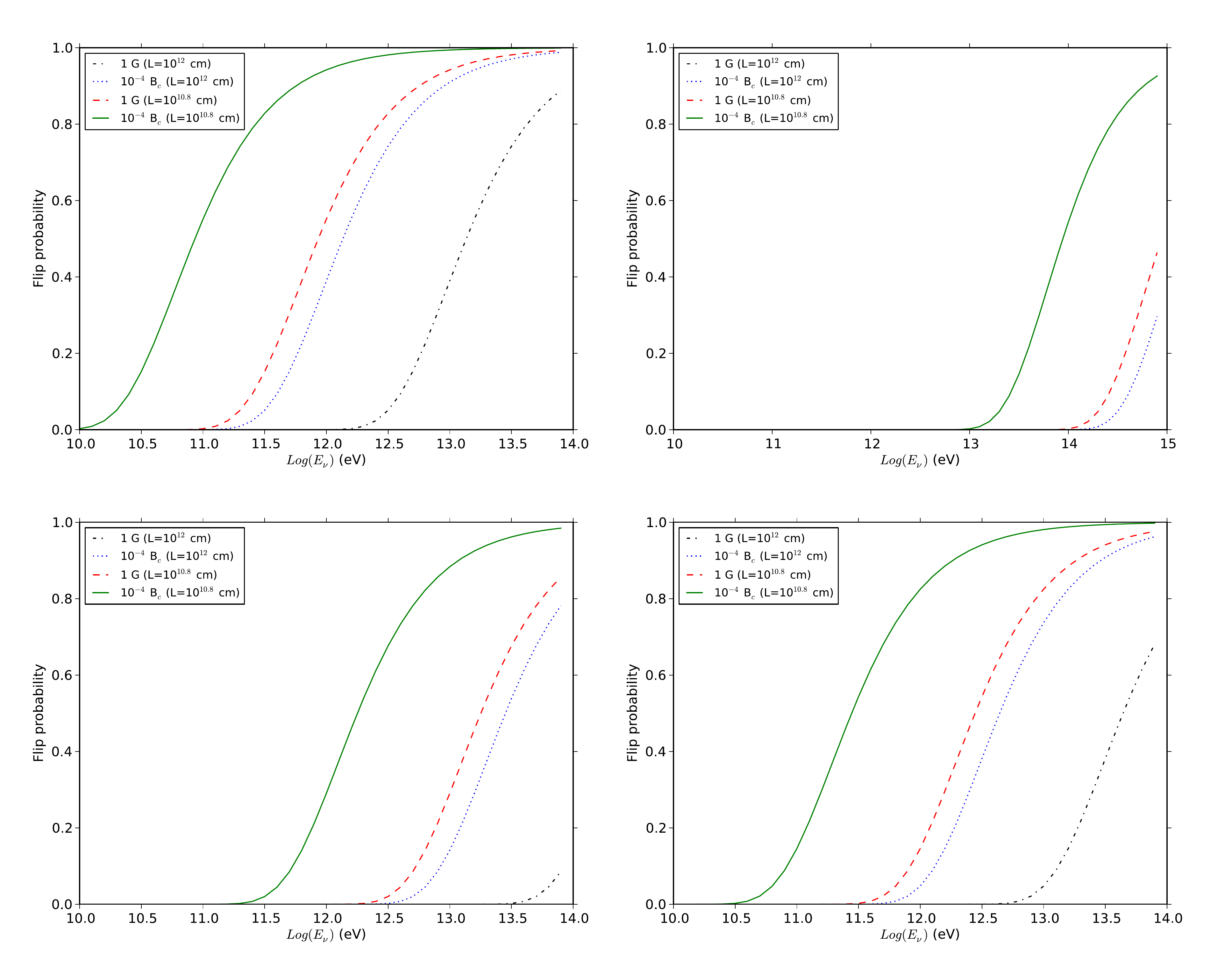}
\caption{The flip probability is plotted as a function of neutrino energy for a strength of magnetic field in the moderate (B=$10^{-4}\,B_c$) and weak (B=1 G) regime and at a distance of $10^{10.8}$ cm and $10^{12}$ cm. We have used the best-fit values of the two-neutrino mixing  (solar, top left; atmospheric, top right; and accelerator, bottom left) and three-neutrino mixing (bottom right). \label{flip}}
\end{figure*}

\appendix
\section{Effective Potential}
The functions of the neutrino effective potential at moderate magnetic field limit are 
\bary
F_m&=&\biggl(1+2\frac{E^2_\nu}{m^2_W}\biggr)K_1(\sigma_l)+2\sum^\infty_{n=1}\lambda_n\biggl(1+\frac{E^2_\nu}{m^2_W}  \biggr)K_1(\sigma_l\lambda_n)\nonumber\\
G_m&=&\biggl(1-2\frac{E^2_\nu}{m^2_W}\biggr)K_1(\sigma_l)-2\sum^\infty_{n=1}\lambda_n\frac{E^2_\nu}{m^2_W} K_1(\sigma_l\lambda_n)\nonumber\\
J_m&=& \frac34 K_0(\sigma_l)+\frac{K_1(\sigma_l)}{\sigma_l}+\sum^\infty_{n=1}\lambda^2_n\biggl[K_0(\sigma_l\lambda)+\frac{K_1(\sigma_l\lambda)}{\sigma_l\lambda}\nonumber\\
&&\hspace{5.2cm}-\frac{K_0(\sigma_l\lambda)}{2\lambda^2_n}  \biggr]\nonumber\\
H_m&=& \frac{K_1(\sigma_l)}{\sigma_l}+ \sum^\infty_{n=1}\lambda^2_n   \biggl[\frac{K_1(\sigma_l\lambda)}{\sigma_l\lambda} - \frac{K_0(\sigma_l\lambda)}{2\lambda^2_n}  \biggr]
\eary
and at weak magnetic field limit are
\bary
F_w&=&\biggl(2+2\frac{E^2_\nu}{m^2_W}\biggr) \biggl(\frac{K_0(\sigma_l)}{\sigma_l}+2\frac{K_1(\sigma_l)}{\sigma_l^2} \biggr) \frac{B_c}{B}-K_1(\sigma_l)\nonumber\\
G_w&=&K_1(\sigma_l)-\frac{2B_c}{B}\frac{E^2_\nu}{m^2_W}    \biggl(\frac{K_0(\sigma_l)}{\sigma_l}+2\frac{K_1(\sigma_l)}{\sigma_l^2}\biggr) \nonumber\\
J_w&=&\biggl(\frac12+\frac{3B_c}{B\,\sigma_l^2}\biggr)K_0(\sigma_l)+\frac{B_c}{B}  \biggl(1+\frac{6}{\sigma_l^2}\biggr)     \frac{K_1(\sigma_l)}{\sigma_l}\nonumber\\
H_w&=& \biggl(\frac12+\frac{B_c}{B\,\sigma_l^2} \biggr)K_0(\sigma_l)+\frac{B}{B_c} \biggl(\frac{2}{\sigma_l^2}-\frac12\biggr)\frac{K_1(\sigma_l)}{\sigma_l}\nonumber\\
\eary
where $\lambda^2_n=1+2\,n\,B/B_c $, K$_i$ is the modified Bessel function of integral order i, $\alpha_l=\beta\mu(l+1)$ and $\sigma_l=\beta m_e(l+1)$.\\
\section{Probabilities}
The flavor ratio at the internal shocks and the envelope of the star are
\be
{\pmatrix
{
\nu_e   \cr
\nu_\mu   \cr
\nu_\tau   \cr
}_{is}}
=
{\pmatrix
{
P_{ee,is}	  & P_{em,is}	  & P_{et,is}\cr
P_{me,is}	  & P_{mm,is}	  & P_{mt,is}\cr
P_{te,is}	  & P_{tm,is}	  & P_{tt,is}\cr
}}
{\pmatrix
{
\nu_e   \cr
\nu_\mu   \cr
\nu_\tau   \cr
}_{c}}
\label{matrixosc}
\ee
and
\be
{\pmatrix
{
\nu_e   \cr
\nu_\mu   \cr
\nu_\tau   \cr
}_{ss}}
=
{\pmatrix
{
P_{ee,ss}	  & P_{em,ss}	  & P_{et,ss}\cr
P_{me,ss}	  & P_{mm,ss}	  & P_{mt,ss}\cr
P_{te,ss}	  & P_{tm,ss}	  & P_{tt,ss}\cr
}}
{\pmatrix
{
\nu_e   \cr
\nu_\mu   \cr
\nu_\tau   \cr
}_{is}}
\label{matrixosc}
\ee
respectively Here {\rm ss} and {\rm is} are the envelope of star and internal shocks, respectively.  The different neutrino probabilities can be written as \citep{gon03,gon08}
{\scriptsize 
\bary
P_{ee}&=&1-4s^2_{13,m}c^2_{13,m}S_{31}\,,\nonumber\\
P_{\mu\mu}&=&1-4s^2_{13,m}c^2_{13,m}s^4_{23}S_{31}-4s^2_{13,m}s^2_{23}c^2_{23}S_{21}-4
c^2_{13,m}s^2_{23}c^2_{23}S_{32}\,,\nonumber\\
P_{\tau\tau}&=&1-4s^2_{13,m}c^2_{13,m}c^4_{23}S_{31}-4s^2_{13,m}s^2_{23}c^2_{23}S_{21}-4
c^2_{13,m}s^2_{23}c^2_{23}S_{32}\,,\nonumber\\
P_{e\mu}&=&4s^2_{13,m}c^2_{13,m}s^2_{23}S_{31}\,,\nonumber\\
P_{e\tau}&=&4s^2_{13,m}c^2_{13,m}c^2_{23}S_{31}\,,\nonumber\\
P_{\mu\tau}&=&-4s^2_{13,m}c^2_{13,m}s^2_{23}c^2_{23}S_{31}+4s^2_{13,m}s^2_{23}c^2_{23}S_{21}+4
c^2_{13,m}s^2_{23}c^2_{23}S_{32}\,,\nonumber\\
\eary
}
where
\be
\sin
2\theta_{13,m}=\frac{\sin2\theta_{13}}{\sqrt{(\cos2\theta_{13}-2E_{\nu}V_{eff,k}/\delta
    m^2_{32})^2+(\sin2\theta_{13})^2}},
\ee
and
\be
S_{ij}=\sin^2\biggl(\frac{\Delta\mu^2_{ij}}{4E_{\nu}}L\biggr).
\ee
Here $\Delta\mu^2_{ij}$ are given by 
\bary
\Delta\mu^2_{21}&=&\frac{\Delta
  m^2_{32}}{2}\biggl(\frac{\sin2\theta_{13}}{\sin2\theta_{13,m}}-1\biggr)-E_{\nu}V_{eff,k}\,,\nonumber\\
\Delta\mu^2_{32}&=&\frac{\Delta
  m^2_{32}}{2}\biggl(\frac{\sin2\theta_{13}}{\sin2\theta_{13,m}}+1\biggr)+E_{\nu}V_{eff,k}\,,\nonumber\\
\Delta\mu^2_{31}&=&\Delta m^2_{32} \frac{\sin2\theta_{13}}{\sin2\theta_{13,m}}\,,
\eary
where
\bary
\sin^2\theta_{13,m}&=&\frac12\biggl(1-\sqrt{1-\sin^22\theta_{13,m}}\biggr)\,,\nonumber\\
\cos^2\theta_{13,m}&=&\frac12\biggl(1+\sqrt{1-\sin^22\theta_{13,m}}\biggr)\,.
\eary
where the neutrino effective potentials  $V_{eff,k}$ are given in section 3.  In vacuum, the flavor ratio (between  the surface of the star and the Earth) is affected by the oscillation probabilities \citep{2007fnpa.book.....G, 1989neas.book.....B}
\bary
P_{\nu_\alpha\to\nu_\beta} &=&\mid <  \nu_\beta(t) | \nu_\alpha(t=0) >  \mid\cr
&=&\delta_{\alpha\beta}-4 \sum_{j>i}\,U_{\alpha i}U_{\beta i}U_{\alpha j}U_{\beta i}\,\sin^2\biggl(\frac{\delta m^2_{ij} L}{4\, E_\nu}   \biggr)\,.
\eary
where  the neutrino  mixing matrix  $U_{ij}$ is given by \cite{gon03,akh04,gon11}
\be\label{mixing}
U =
{\pmatrix
{
c_{13}c_{12}                    & s_{12}c_{13}                    & s_{13}\cr
-s_{12}c_{23}-s_{23}s_{13}c_{12} & c_{23}c_{12}-s_{23}s_{13}s_{12}   & s_{23}c_{13}\cr
s_{23}s_{12}-s_{13}c_{23}c_{12}  &-s_{23}c_{12}-s_{13}s_{12}c_{23}   &  c_{23}c_{13}.\cr
}},
\ee
Here $s_{ij}=\sin\theta_{ij}$ and  $c_{ij}=\cos\theta_{ij}$ and we have taken the Dirac phase $\delta=0$.   Taking into account the effect of internal shocks, envelope of star and the vacuum, the probabilities are given by
\bary
P^*_{11}&=& 0.82\,P_{11}+0.55\,P_{21}+0.19\,P_{31} \nonumber\\
P^*_{12}&=& 0.82\,P_{12}+0.55\,P_{22}+0.19\,P_{32} \nonumber\\
P^*_{13}&=& 0.82\,P_{13}+0.55\,P_{23}+0.19\,P_{33} \nonumber\\
P^*_{21}&=& -0.51\,P_{11}+0.51\, P_{21}+ 0.69\,P_{31}\nonumber\\
P^*_{22}&=& -0.51\,P_{12}+0.51\,P_{22}+0.69\,P_{32}\nonumber\\
P^*_{23}&=& -0.51\,P_{13}+0.51\,P_{23}+0.69\,P_{33}\nonumber\\
P^*_{31}&=& 0.28\,P_{11}+-0.66\,P_{31}+0.69\,P_{31} \nonumber\\
P^*_{32}&=& 0.28\,P_{12}+-0.66\,P_{22}+0.69\,P_{32} \nonumber\\
P^*_{33}&=& 0.28\,P_{13}+-0.66\,P_{23}+0.69\,P_{33} 
\eary
where 
\bary
P_{11}&=&P_{ee,ss}P_{ee,is}+P_{e\mu,ss}P_{\mu e,is}+P_{e\tau,ss}P_{\tau e,is}\nonumber \\
P_{12}&=&P_{ee,ss}P_{e\mu,is}+P_{e\mu,ss}P_{\mu\mu,is}+P_{e\tau,ss}P_{\tau\mu,is}\nonumber\\
P_{13}&=&P_{ee,ss}P_{e\tau,is}+P_{e\mu,ss}P_{\mu\tau,is}+P_{e\tau,ss}P_{\tau\tau,is}\nonumber\\
P_{21}&=&P_{e\mu,ss}P_{ee,is}+P_{\mu\mu,ss}P_{\mu e,is}+P_{m\tau,ss}P_{\tau e,is}\nonumber\\
P_{22}&=&P_{\mu e,ss}P_{e\mu,is}+P_{\mu\mu,ss}P_{\mu \mu,is}+P_{m\tau,ss}P_{\tau \mu,is}\nonumber\\
P_{23}&=&P_{\mu e,ss}P_{e\tau,is}+P_{\mu\mu,ss}P_{\mu \tau,is}+P_{m\tau,ss}P_{\tau \tau,is}\nonumber\\
P_{31}&=&P_{e\tau,ss}P_{ee,is}+P_{\tau\mu,ss}P_{\mu e,is}+P_{\tau\tau,ss}P_{\tau e,is}\nonumber\\
P_{32}&=&P_{\tau e,ss}P_{e\mu,is}+P_{\tau\mu,ss}P_{\mu \mu,is}+P_{\tau\tau,ss}P_{\tau \mu,is}\nonumber\\
P_{33}&=&P_{\tau e,ss}P_{e\tau,is}+P_{\tau\mu,ss}P_{\mu \tau,is}+P_{\tau\tau,ss}P_{\tau \tau,is}
\eary

\end{document}